\documentclass{eas}
\usepackage[utf8]{inputenc}
\usepackage[english]{babel}
\usepackage[T1]{fontenc}
\usepackage{amsmath, amsfonts, amssymb}
\usepackage[dvips]{graphicx}
\usepackage{subfig}
\usepackage[dvipsnames]{pstricks} 
\usepackage{pstricks-add} 
\usepackage[authoryear]{natbib}
\usepackage{hyperref}
\hypersetup{%
            unicode=True,%
            colorlinks=True,%
            pdfauthor={J.~Morin},%
            pdftitle={Magnetic fields from very low mass stars to brown dwarfs},%
            pdfcreator={\LaTeXe}%
}
\bibpunct{(}{)}{;}{a}{}{,}
\usepackage{journal}
\def\mstar{\hbox{$M_{\star}$}}

\def\msun{\hbox{${\rm M}_{\odot}$}}

\def\vsini{\hbox{$v\sin i$}}

\def\kms{\hbox{km\,s$^{-1}$}}
\def\um{\hbox{$\mu$m}}
\def\Prot{\hbox{$P_{\rm rot}$}}
\def\d{\hbox{$\rm d$}}

\def\Bf{\hbox{$B\!f$}}
\def\kG{\hbox{$\rm kG$}}
\def\degr{\hbox{$^\circ$}}

\def\Ha{\hbox{${\rm H}\alpha$}}

\renewcommand*{\vec}[1]{\mbox{$\mathbf{\displaystyle#1}$}}

\newcommand{\EE}[1]{\times 10^{#1}}
\newcommand{\avg}[1]{\left\langle #1 \right\rangle}

\def\ea{\hbox{\itshape et al.}}
\def\eg{\hbox{\itshape e.g., }}

\def\enumi{\hbox{\itshape (i)}}
\def\enumii{\hbox{\itshape (ii)}}
\def\enumiii{\hbox{\itshape (iii)}}
\def\enumiv{\hbox{\itshape (iv)}}

\begin{document}

\title{Magnetic fields from low mass stars to brown dwarfs}
\author{J.~Morin}%
  \address{%
    Dublin Institute for Advanced Studies, School of Cosmic Physics, 31
    Fitzwilliam Place, Dublin 2, Ireland
  }%
  \secondaddress{%
    Institut für Astrophysik, Georg-August-Universität, Friedrich-Hund-Platz 1,
    D-37077 Göttingen, Germany; %
    \email{jmorin@gwdg.de}
  }

\begin{abstract}
Magnetic fields have been detected on stars across the H-R diagram and substellar objects either
directly by their effect on the formation of spectral lines, or through the activity phenomena they
power which can be observed across a large part of the electromagnetic spectrum. Stars show a very
wide variety of magnetic properties in terms of strength, geometry or variability. Cool stars
generate their magnetic fields by dynamo effect, and their properties appear to correlate --- to
some extent --- with stellar parameters such as mass, rotation and age.
With the improvements of instrumentation and data analysis techniques, magnetic fields can now be
detected and studied down to the domain of very-low-mass stars and brown dwarfs, triggering
new theoretical works aimed, in particular, at modelling dynamo action in these objects. After a
brief discussion on the importance of magnetic field in stellar physics, the basics of dynamo theory
and magnetic field measurements are presented. The main results stemming from observational and
theoretical studies of magnetism are then detailed in two parts: the fully-convective transition,
and the very-low mass stars and brown dwarfs domain.
\end{abstract}

\maketitle

\section{Introduction}
Magnetic fields have often been neglected in early astrophysical theories, both for simplicity and
lack of observational data. However, recent advances point out the crucial role of magnetic fields
in many aspects of stellar physics; besides the number of stellar magnetic field measurements is
constantly growing and the coverage of the H-R diagram constantly improving \cite[][]{Donati09,
Reiners12b}. Magnetic fields are particularly important during the formation and early evolution of
stars: for instance they can oppose the gravitational collapse of molecular clouds (see chapter by
P.~Hennebelle, this book), the launching and confinement of bipolar jets observed on protostars are
MHD phenomena \cite[][]{Tsinganos09}, and matter accreted from the circumstellar disc onto a T Tauri
star is thought to be channelled along magnetic fields lines \cite[][]{Bouvier07}.
Magnetic fields are also know to play a key role in the rotational evolution of low-mass stars,
indeed most of the phenomena which drive this evolution (\eg star-disc coupling, stellar winds)
involve magnetic fields \cite[][]{Bouvier09}. An additional issue that has emerged since the
discovery of the first exoplanet \cite[][]{Mayor95} has been the impact stellar of magnetic fields
and activity on the formation, evolution, and habitability of planetary systems, as well as on the
detection of planets orbiting around active stars (see chapter by X.~Bonfils, this book). All these
topics are quite relevant to low mass stars and brown dwarfs, as demonstrated by recent
studies such as the detection of outflows on substellar objects \cite[][]{Whelan05}, the
observational and theoretical works addressing the rotational evolution of M dwarfs \cite[][]{Irwin11,
Reiners12a}, and the ongoing or planned planet-search programs targeting M dwarfs
\cite[][]{Bonfils11}. Moreover, since some of them host very strong magnetic fields, low mass stars
and brown dwarfs are particularly interesting laboratories to study the effect magnetism on stellar
internal structure \cite[][]{Chabrier07}. Last but not least, low mass stars and brown dwarfs
appear as a vital link to build a consistent picture of dynamo action in astrophysical bodies from
planets to solar-type stars \cite[][]{Christensen09}.

\section{Dynamo action in cool stars}
\label{sec:cs}
A fundamental difference exists among stars regarding the origin of the magnetic fields that are
detected in their outer atmosphere. On the one hand, in some stars with outer radiative zones or in
compact objects, simple and steady magnetic fields are observed. They are generally thought to be
fossil fields\footnote{Although the possibility for magnetic field generation in stellar radiative
zones is a debated topic \cite[\eg][]{Braithwaite04}}, \ie\ to be the remnant of a field generated
at an earlier phase of their evolution.
On the other hand, cool stars (\ie\ spectral types later than mid-F) as well as brown dwarfs possess
an outer convective envelope where the magnetic diffusivity is strongly enhanced by fluid motions
\cite[\eg][]{Ruediger11}, resulting in field decay times of the order of a decade. The very
existence of significant magnetic fields on cool stars with ages of the order of a Gyr, as well as
their dynamic nature (in particular the existence of magnetic cycles) point out that magnetic fields
have to be generated in these objects, namely by dynamo action.
\begin{figure}
  \centering\includegraphics[width=0.7\textwidth]{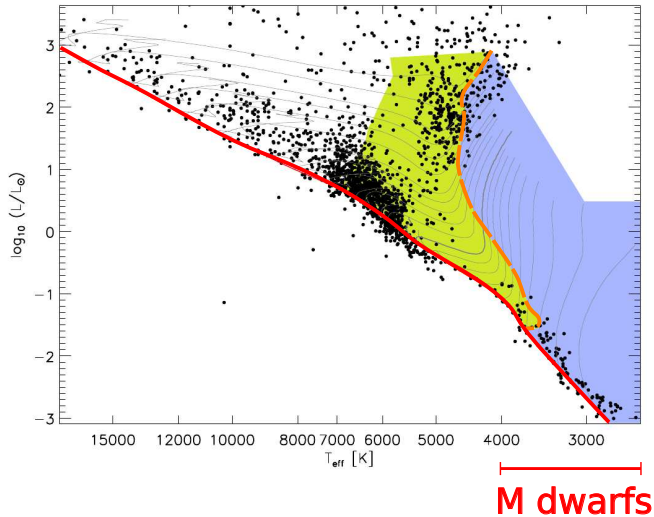}
  \caption{Hertzsprung--Russel diagram with evolutionary tracks from \cite{Siess00}. Cool
stars are located in the green (partly convective) and blue (fully
convective) regions. The main sequence is depicted as a red solid line and the fully
convective limit as an orange dashed line. M dwarfs are located at the cool end of the main
sequence. Even cooler L, T and Y dwarfs (\ie\ brown dwarfs) are not represented here. Adapted from
\cite{Reiners08}.}
  \label{fig:convection-hrd}
\end{figure}

\subsection{A few words of history}
\label{sec:cs-hist}
\cite{Hale1908} first measured strong magnetic fields (several kilogauss) in sunspots. This
observation revealed at the same time the existence of the solar magnetic field as well as its
cyclic nature, according with previous knowledge about sunspots: the 11-yr cyclic variation of their
number \cite[][]{Schwabe1844}, and their migration in latitude during the cycle \cite[the
``butterfly diagram'',][]{Maunder1904}. These results were later complemented by the discovery of
the polarity law of sunspots making the magnetic cycle 22~yr long \cite[][]{Hale1919}. In addition
to sunspots, a weak (of the order of 1~G) large-scale component of the solar magnetic field which
follows the the same 22-yr cycle is also observed \cite[][]{Babcock55, Babcock61}. Both
the relative importance of the dipole and quadrupole components, as well as their tilt angle with
respect to the solar rotation axis were shown to evolve during the solar cycle
\cite[][]{Sanderson03}. Although the strong magnetic fields associated with sunspots and the
large-scale solar field are likely the two features that receive the most attention for models of
the solar cycle, a whole ``zoo'' of solar magnetic features exists, the reader is referred to 
\cite{Solanki06}.

\cite{Larmor1919} first proposed that the solar magnetic field
could be induced by motions of an electrically conducting fluid. He also pointed
out the possible role of differential rotation to generate a strong
ordered azimuthal field deep inside the Sun, such as mentioned by
\cite{Hale1919} to explain the properties of sunspots. This was the first step
toward the concept of solar magnetism powered by dynamo action, \ie\ a mechanism
able to convert kinetic energy of plasma motions into magnetic energy in a
self-sustained manner. However this idea was severely hampered by
\cite{Cowling33}'s anti-dynamo theorem which states that a purely axisymmetric
field cannot be sustained by dynamo action. \cite{Elsasser46}\footnote{Elsasser's work was
originally mostly focused on the geodynamo problem, but his formalism was also rapidly used for the
solar dynamo.} then introduced the poloidal--toroidal decomposition of the magnetic field (see
Appendix \ref{sec:pol-tor}) and formalized Larmor's idea on the role of differential rotation by
showing that it could transform a poloidal field into a stronger toroidal one,  this is now
referred
to as the $\Omega$-effect. He also demonstrated that differential
rotation alone was not enough to sustain a dynamo (toroidal theorem), another mechanism is indeed
required to regenerate a poloidal component from the toroidal field. 

A first solution was initiated by \cite{Parker55}: convective motions in the solar envelope are
deflected by the Coriolis force, this twists the field lines in a systematic way and allows the
regeneration of poloidal field from toroidal field. The combination of this effect with the
aforementioned $\Omega$-effect constitutes the first model of an hydromagnetic self-sustained
dynamo. \cite{Steenbeck66} then introduced the mean-field theory and the $\alpha$-effect which
turned out to be closely related to Parker's concept of cyclonic convection.

\subsection{A few words of mean-field theory}
\label{sec:cs-theory}
The dynamo effect consists in generating and sustaining a magnetic field from the motion of an
electrically conducting fluid, \ie\ converting kinetic energy into magnetic energy. Dynamos are
thought to operate in many astrophysical objects from telluric and gaseous planets, to stars and
galaxies. The main concepts as well some vocabulary often encountered in the literature are
presented here. For a more detailed view of dynamo theory, the reader is referred to the
following reviews and books: \cite{Ossendrijver03, Ruediger04, Brandenburg05,
Dormy07, Charbonneau10}.
The fundamental equation for this problem is the induction equation, it can be derived from the
Maxwell--Faraday and Maxwell--Ampère equations combined with the generalized Ohm's law\footnote{As
opposed to Maxwell's equation and Ohm's law, the induction equation does not depend on the choice
of SI or CGS units}:
\begin{equation}
  \label{eq:induction}
  \dfrac{\partial \vec{B}}{\partial t} = %
  \nabla \times \left(\vec{u} \times \vec{B}%
  - \eta \nabla \times \vec{B} \right) = %
  \underbrace{\nabla \times \left(%
  \vec{u} \times \vec{B} \right) }_{induction}%
  + \underbrace{\phantom{\cos}\eta \vec{\Delta B}\phantom{\cos}}_{dissipation}%
  \text{,}
\end{equation}
where $\vec{B}$ and $\vec{u}$ are respectively the magnetic field and fluid
velocity vectors, and $\eta$ 
is the magnetic diffusivity. The ratio between the two terms of RHS of the induction equation can be
approximated by the magnetic Reynolds number:
\begin{equation}
  {\rm Rm} = \dfrac{\rm induction}{\rm
diffusion}=\dfrac{\frac{u_0 B_0}{L}}{\frac{\eta B_0}{L^2}}=\dfrac{u_0\,L}{\eta}\text{,}
\end{equation}
where $u_0$, $B_0$ and $L$ are characteristic velocity, magnetic field modulus and length scale.
For dynamo action to be possible $\rm Rm$ must therefore necessarily be larger than unity, although
this condition is not sufficient. We note that in the convection zones of cool stars 
estimates show that this condition is satisfied \cite[\eg][table~1]{Brandenburg05}.
Solar and stellar physicists are generally interested in large-scale dynamos,
\ie\ which generate magnetic field on spatial scales larger than that of the convection and can
explain the properties of magnetic cycles\footnote{Although a turbulent or small-scale dynamo likely
coexists \cite[][]{Durney93, Dorch02,Voegler07}}. Not any flow can act as a large-scale dynamo, in
particular a number of anti-dynamo theorem (\cf\ Sec.~\ref{sec:cs-hist}) indicate that some specific
properties are required.

In order to fully address the dynamo problem, one has to consider, in addition to the induction
equation, the $\nabla\cdot\vec{B}=0$ Maxwell's equation, as well as an equation of state and the
Navier-Stokes equations to describe the fluid motion. We also note that according to the anti-dynamo
theorems, the dynamo problem is intrinsically 3-D and cannot be addressed by considering 2-D flow
and magnetic field. The mean field approach has been developed by \cite{Steenbeck66, Steenbeck69a,
Steenbeck69b}\footnote{These papers were originally published in German and later translated in
English by \cite{Moffatt70, Roberts71}} in order to allow for a simplified treatment of the dynamo
problem, this theoretical framework remains important in our understanding of the solar dynamo as
well as in the vocabulary. This is a kinetic approach, \ie\ the induction equation is solved for a
given flow, the feedback of the Lorentz force on the fluid motion is not considered. Its specificity
consists in decomposing both the magnetic and velocity field into the sum of a mean and a
fluctuating components, generally understood as the axisymmetric and non-axisymmetric parts, though
any averaging operator for which the Reynolds rules apply can be used \cite[see \eg][]{Raedler07b}.
Let $\avg{\cdot}$ be this averaging operator: 
\begin{equation}
  \label{eq:mf-decomp}
  \vec{x} = \avg{\vec{x}} +\vec{x}'\text{,}
\end{equation}
where $\vec{x}$ can be either $\vec{u}$ or $\vec{B}$, and $\vec{x}'$ is the
fluctuating component. By introducing this decomposition into equation
(\ref{eq:induction}), one gets:
\begin{equation}
  \label{eq:mf-induction}
  \dfrac{\partial \avg{\vec{B}}}{\partial t} = \nabla \times \left(%
  \avg{\vec{u}} \times \avg{\vec{B}} + %
  {\avg{\vec{u}' \times \vec{B}'}}%
  - \eta \nabla \times \avg{\vec{B}}
  \right)\text{.}
\end{equation}
The main difference with (\ref{eq:induction}) is the presence of the term
$\avg{\vec{u}' \times \vec{B}'}$ which is referred to as the turbulent electromotive force. It is
the presence of the fluctuating fields in the source term of the induction equation
for the mean magnetic field which can circumvent Cowling's theorem.
Up to this point no approximation has been made, but in order to solve Eq.~(\ref{eq:mf-induction}),
a closure equation linking the fluctuating quantities to the mean magnetic field is
required, this is detailed  by \eg \cite{Raedler07b}. The simplest approach assumes
\enumi\ a clear scale separation between and the mean and fluctuating components and \enumii\ that
the fluctuating velocity field corresponds to homogeneous and isotropic turbulence, the turbulent
EMF is then expressed as:
%
\begin{equation}
  \avg{\vec{u}' \times \vec{B}'} = \alpha \avg{\vec{B}} - \beta \nabla \times
  \avg{\vec{B}}\text{.}
\end{equation}
By substituting this expression for the turbulent EMF and introducing the
poloidal-toroidal decomposition (\ref{eq:pol-tor-pot}) into
equation (\ref{eq:mf-induction}) one gets the evolution equations for the mean poloidal and
toroidal field respectively:
\begin{equation}
  \label{eq:mf-induction-pol-tor}
  \left\lbrace
  \begin{aligned}
    \dfrac{\partial A}{\partial t} + %
      \dfrac{1}{s} %
      (\vec{u_{\rm pol}}\cdot\nabla)(s A) & = %
    \phantom{s \vec{B_{\rm pol}} \cdot \nabla\Omega -} %
      \phantom{-}\alpha B \phantom{\alpha_1^2} \!+\;%
      (\eta+\beta)  \nabla_1^2 A %
      \\
    \dfrac{\partial B}{\partial t} + %
      s (\vec{u_{\rm pol}}\cdot\nabla)%
      (\dfrac{B}{s})\phantom{s} & = %
    \underbrace{s \vec{B_{\rm pol}} \cdot \nabla\Omega}_%
      {\text{$\Omega$-effect}} %
      \; - %
      \underbrace{\alpha \nabla_1^2 A}_%
      {\text{$\alpha$-effect}} + \; %
      \underbrace{(\eta+\beta) \nabla_1^2  B}_%
      {dissipation} \text{,}
      \\
  \end{aligned}
  \right.
\end{equation}
\begin{equation}
  \text{with }s=r\sin{\theta}\;\text{;}\; %
  \nabla_1^2=\left(\Delta-\dfrac{1}{s^2}\right)\;\text{;}\; %
  \vec{B_{\rm pol}=\nabla\times(A\vec{e_\phi})}\text{,}
\end{equation}
where spherical coordinates $(r,\theta,\phi)$ are used and $\vec{B_{\rm pol}}$ is the poloidal
component of the mean magnetic field. A few comments can be made on equation
(\ref{eq:mf-induction-pol-tor}):
\begin{itemize}
  \item The $\alpha$-effect is related to the mean kinetic helicity
of the turbulent flow\footnote{The kinetic helicity
$\avg{\vec{u}\cdot(\nabla\times\vec{u})}$ characterize the fact that the fluid
has both a rotation movement and a translation collinear with the rotation axis,
as would be the case for a movement along a helix.}, thus representing in a
more formalized way Parker's idea of convective motions systematically deflected and twisted
by the Coriolis force. 
  \item The $\beta$-effect on its side is related to the kinetic energy of the turbulent flow and
acts as an enhanced dissipation. In stellar convection zones, estimates show that this turbulent
diffusivity is higher than the intrinsic diffusivity of the plasma by several order of magnitudes.
  \item The $\Omega$-effect can only produce toroidal field from a poloidal component, and therefore
a dynamo cannot rely on this effect alone, the alpha effect is necessary to regenerate a poloidal
component of the field. On the opposite, as the $\alpha$-effect is present as a source term in both
the poloidal and the toroidal induction equations, a dynamo can rely on this effect
alone.
  \item The $\Omega$-effect depends on differential rotation, solid-body rotation is related to the
$\alpha$-effect through the Coriolis force that generates kinetic helicity.
  \item Dynamos in which toroidal field generation is predominantly due to the
$\alpha$-effect ($\Omega$) are called $\alpha^2$ ($\alpha\Omega$), and the
intermediate case where both terms have similar order of magnitude is sometimes
referred to as $\alpha^2\Omega$. 
\end{itemize}

\subsection{Later developments and the role of the tachocline}
\label{sec:cs-theory}
An important success of the mean-field theory with parametrized $\alpha$-effect resides in that
$\alpha\Omega$ dynamos can exhibit cyclic solutions that can be related to the solar cycle
\cite[][]{Steenbeck69a}. 
However, additional issues have emerged which have questioned the mean-field approach and the
validity of the results obtained, for a critical detailed critical view see \cite{Spruit11}. First,
the generation of the very strong toroidal fields required
to explain the observed properties of sunspots appeared incompatible with the magnetic buoyancy
instability \cite[][]{Parker75}.
Secondly,  the mechanism that generates poloidal magnetic field from a toroidal
component is still debated, and alternatives to and improvements of the $\alpha$-effect are
proposed, such as the Babcock-Leighton mechanism \cite[][]{Babcock61,Leighton69}, or
generation mechanisms based on MHD instabilities \cite[see the discussion on poloidal field
generation mechanisms in][]{Charbonneau10}.
Third, there are intrinsic limitations of the mean-field kinematic approach itself which
does not consider the feedback of the magnetic on the fluid through the Lorentz force. 
Finally, a major change came from the field of helioseismology,  the measurements of the
internal rotation profile of the Sun dramatically differed from theoretical expectations based on
mean-field modelling. Even more crucially these measurements revealed
the existence of the tachocline, a thin layer of strong shear located at the interface between the
inner radiative core and the outer convective zone. 

A number of subsequent studies have tried to overcome these issues in the framework of mean-field
$\alpha\Omega$ modelling. In particular, flux transport dynamo models -- based on a
Babcock-Leighton generation mechanism -- attribute an important role to meridional
circulation \cite[][]{Choudhuri95,Dikpati99}; on their side interface dynamo models assume that the
$\alpha$- and $\Omega$-effects are spatially segregated on either sides of the core-envelope
interface \cite[][]{Parker93,Charbonneau97}. These studies attribute a crucial to the solar
tachocline as being a place where toroidal fields can be strongly amplified up to the field strength
required to explain sunspot properties and on timescales compatible with the solar cycle length. In
parallel, the constantly-increasing computing capabilities have allowed to perform direct numerical
simulations of dynamo action in spherical shells at high resolution.
These simulations resolve the full set of equations that self-consistently describe the temporal
evolution of the velocity and magnetic field, but are  limited to regimes of parameters quite remote
from real astrophysical objects. With numerical simulations, the interplay between
differential rotation and magnetic fields can be studied \cite[\eg][]{Brown10}; the presence of a
stratified layer that mimics the tachocline appears to facilitate the building of strong large-scale
toroidal fields \cite[][]{Browning06} though it is not essential
\cite[][]{Brown10}; and cyclic polarity reversal start being observed \cite[][]{Kapyla10, Brown11}.
It is also worth noting that mean-field formalism remains a very useful tool to understand the
magnetic field generation in numerical simulation \cite[][]{Schrinner07}.

\section{Direct methods for magnetic field measurements}

\subsection{Zeeman effect}
\label{sec:measurements-zeeman}
\begin{figure}
  \centering %
  \includegraphics[width=.9\textwidth]{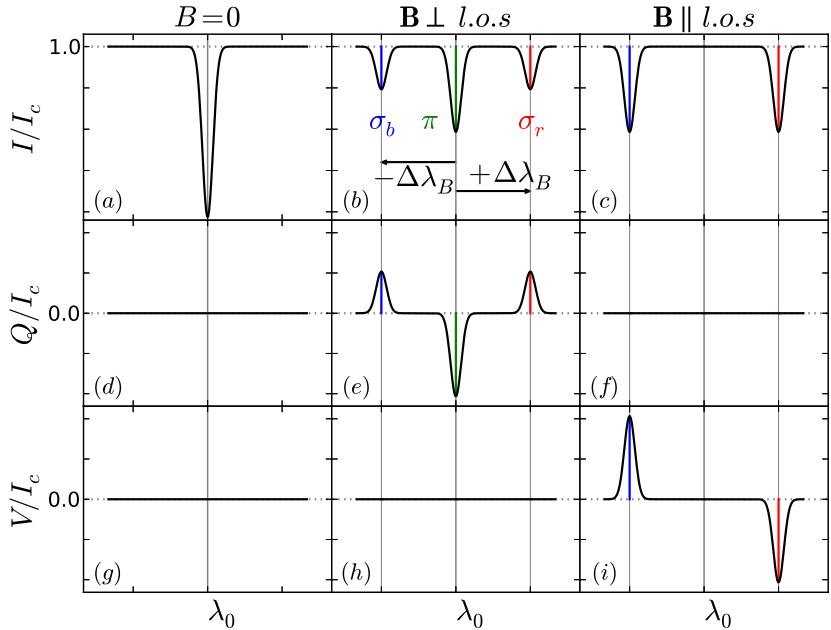}
  \caption{Schematic view of the Zeeman effect for a normal Zeeman triplet and a large splitting.
The rows show from top to bottom the Stokes  parameters I (intensity), Q (linear polarisation) and V
(circular polarisation) normalized to the  unpolarized continuum I$_c$ (see
Appendix~\ref{sec:stokes-param}). Columns show from left to right the null-field  case, and two
magnetic cases with $\vec{B}$ perpendicular and parallel to the observer's line of sight. In the
middle and right columns, the $\pi$, $\sigma_b$ and $\sigma_r$ components are respectively marked
with a green, blue and red vertical line. The scale is the same for all panels. The x-axis ticks
are in units of Zeeman splitting $\Delta\lambda_B$ (Eq.~\ref{eq:zeeman-broadening}), and $\lambda_0$
is the central wavelength of the line without a magnetic field. The reference level (1 for I, 0 for
Q, V) is plotted as a dotted gray line in each panel. The actual sign of Q and V depends on the
polarity of the field.}
  \label{fig:tech-zeeman-sketch}
\end{figure}

Direct measurements of stellar magnetic fields at the photospheric level rely
on the properties of the Zeeman effect. This effect of the presence of a
magnetic field on the formation of spectral lines has two aspects, 
represented on Fig.~\ref{fig:tech-zeeman-sketch}:
\begin{itemize}
  \item A single spectral line splits into several components. In the
case of the so-called normal Zeeman triplet three components are observable: a
$\pi$ component lying at the same wavelength as the null-field line, and two
$\sigma$ components shifted towards the red and the blue by an equal amount.
The amount of $\pi$-to-$\sigma$ splitting in a given spectral line
is proportional to the modulus of the magnetic field $B$, in the CGS unit system it is given by:
\begin{equation}
  \Delta\lambda_B %
                  = \dfrac{\lambda_0^2 e \, g_{\rm eff} B}{4 \pi m_e c^2} %
                  = 4.67 \times 10^{-12}\, \lambda_0^2\, g_{\rm eff} B\text{,}
  \label{eq:zeeman-broadening}
\end{equation}
where $B$ is expressed in Gauss, $\lambda_0$ is the central wavelength of the line without magnetic
field (in nm), and the effective Landé factor $g_{\rm eff}$ is a dimensionless
number that quantities the sensitivity of a given line to the Zeeman effect. Spectral lines which
are considered ``magnetically sensitive'' have Landé factors of the order of 1, up to a few units.
  \item The three components of the Zeeman-split spectral line have different
polarization properties, the observed polarization  depends on the relative
orientation of the vector magnetic field with the line-of-sight of the observer. Stokes V
(circular polarization) is sensitive to the line-of-sight component of the field
(longitudinal field), whereas Q and U (linear) are sensitive to
the component lying in the plane perpendicular to it.
\end{itemize}
 
\begin{figure}
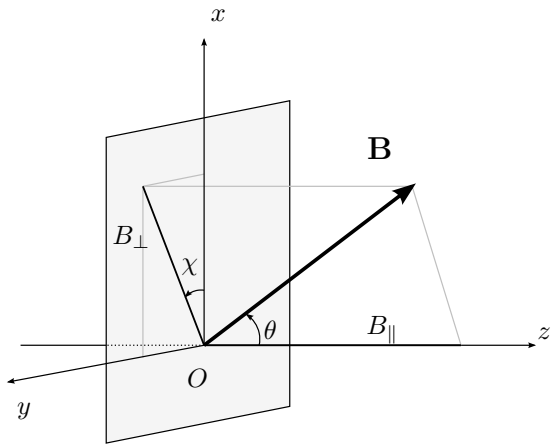

  \centering %
  \include{pst/zeeman-angles}
  \caption{Geometry of the magnetic field vector with respect to the observer's line of sight
  (LoS) for the description of the Stokes parameters. The LoS is along the axis $Oz$. The
  plane $xOy$ is perpendicular to it, with $Ox$ the reference direction of the linear polarisation
  (see Appendix~\ref{sec:stokes-param}). Measurements in circular polarisation (Stokes~V) are
  sensitive to the longitudinal  magnetic field $B_{\parallel}=B\cos{\theta}$.
  Measurements in linear polarisation (Q and U) are sensitive to the transversal component
  $B_{\perp}=B\sin{\theta}$.}     
  \label{fig:tech-zeeman-angles}
\end{figure}

In principle, from a measurement of all four Stokes parameters one can recover the magnetic field
vector. Such an inversion procedure requires to solve the equations of polarised radiative
transfer. The simplest solution is the so-called weak-field approximation which is valid if the
Zeeman splitting $\Delta\lambda_B$ is small compared to the Doppler broadening of the line, in this
case the individual $\pi$ and $\sigma$ components are not resolved, and in unpolarized light the
Zeeman effect results in line broadening rather than splitting. In this case polarisation in
spectral line is a 1$^{\rm st}$ (circular) or 2$^{\rm nd}$ (linear) order effect and the
corresponding amplitudes are small compared to the unpolarised line depth:
\begin{equation}
\left\lbrace
\begin{array}{l}
 Q(\lambda) = -\dfrac{1}{4} g_{\rm eff}^2 \Delta\lambda_B^2 \sin^2\theta
\cos{2\chi} \dfrac{d^2 I_0}{d\lambda^2} \\
U(\lambda) =  -\dfrac{1}{4} g_{\rm eff}^2 \Delta\lambda_B^2 \sin^2\theta
\sin{2\chi} \dfrac{d^2 I_0}{d\lambda^2}  \\
 V(\lambda) = -g_{\rm eff}\Delta\lambda_B \cos{\theta}\dfrac{dI_0}{d\lambda}
\end{array}
\right.
\text{,}
\label{eq:zeeman-weakfield}
\end{equation}
where $I_0$ is the corresponding Stokes I line profile without a magnetic field and the angles
$\theta$ and $\chi$ are defined in Fig.~\ref{fig:tech-zeeman-angles}. A more accurate description
can be obtained with the analytical Unno-Rachkovsky solution \cite[][]{Unno56, Rachkovsky69} or by
numerical solving. For a more detailed view of the  Zeeman effect and other polarizing mechanisms
in astrophysics, as well as an introduction to polarized radiative transfer, and
spectropolarimetry the reader is referred to \cite{Landi92}, \cite{delToroIniesta03} and 
\cite{Landstreet09a,Landstreet09b,Landstreet09c}. 
%

\subsection{Unpolarized spectroscopy}
\label{sec:tech-unpol}
Since for a given field strength, the Zeeman splitting relative to the line wavelength is
$\frac{\Delta\lambda_B}{\lambda_0}\propto\lambda_0\,g_{\rm eff}$, such measurements are generally
performed in high-Landé factor lines located in the red or even infrared part of stellar spectra.
Let's consider the effect of a magnetic field of 2~\kG\ (typical for a sunspot) in the favourable
case of the Fe~\textsc{i} line at 630.25~nm which has $g_{\rm eff}=2.5$. The resulting splitting is
$\Delta\lambda_B=9.2\EE{-3}$~nm, corresponding to a
spectral resolution of $\frac{\lambda_0}{\Delta\lambda_B}=6.8\EE{5}$ or a
velocity of $4.4~ \kms$. This can be directly observed on sunspots. However, for
unresolved stars, spectra correspond to light integrated over the whole visible disc which likely
features a distribution of magnetic field strengths and velocities. In addition, most active stars
rotate with \vsini\ of a least a few \kms, further ``blurring'' the Zeeman splitting. Therefore the
individual $\pi$ and $\sigma$ components are generally not observable\footnote{Among non-degenerate
stars, resolved $\pi$ and $\sigma$ components of Zeeman-split lines can be observed mostly in
chemically peculiar Ap/Bp stars (which host fields of up to several tens of \kG)
\cite[\eg][]{Mathys97}.}, rather the Zeeman effect generally results in
spectral line broadening. In order to disentangle between the effect of the magnetic field and
other sources of broadening it is necessary to observe spectral lines of the same element with both
low and high Landé factors and/or to observe both an active (to be studied) and an inactive (as a
reference) star as similar as possible (temperature, gravity, metallicity). Most models consider
that magnetic regions with assumed homogeneous field modulus $B$, cover a fraction $f$ of the
stellar surface, \cite[][]{Robinson80,Saar88}; or even a range of local field strengths with each a
different filling factor \cite[][]{Johns-Krull99}. The multiple parameters can generally not be
constrained individually, the measured quantity is termed \Bf, the product of the field modulus
$B$ in magnetic regions (assumed homogeneous) times the corresponding filling factor $f$, this
quantity is sometimes called ``magnetic flux'' (although it is expressed in Gauss and not Maxwell)
and reflects the magnetic field modulus averaged over the visible stellar disc. 

M dwarfs are potentially good targets for magnetic field measurements since a significant fraction
of them is very active (see lecture notes by X.~Bonfils, this book), and their spectral energy
distributions peak in the red or infrared where the
Zeeman effect is stronger. Measurements based on atomic lines
are possible for early and mid M dwarfs, however toward late spectral types
molecular lines becomes more prominent (see
chapter by F.~Allard, this book) and further restrict the number of unblended
atomic lines available
for field measurements. The Wing-Ford band of FeH at 0.99~\um\ is composed of many lines
spanning a range of Landé factors and is present through the whole M spectral type,
making it an ideal candidate for magnetic field investigations in low-mass stars
\cite[][]{Valenti01}. To overcome the lack Landé factor values for these lines,
\cite{Reiners06b} have proposed an approach consisting of fitting a stellar spectra in the FeH
region as a linear combination of the spectrum of an inactive star and a very active
star (corrected for FeH band strength and \vsini) for which the magnetic field modulus
has been derived from the analysis of atomic lines. This method has been successfully applied to tens
of stars spanning the whole M dwarf spectral type, and allowed to derive both \vsini\ and
\Bf\ with respective typical accuracies of 1~\kms\ and 0.5--1~\kG. 
Future progress will likely come from theoretical computation of
the Landé factors \cite[][]{Shulyak10} as well as experimental determinations.

\subsection{Spectropolarimetry}
\begin{figure}
  \begin{center}
  \leavevmode  
  \subfloat[$\varphi=-0.10$]{%
    \begin{minipage}{0.40\textwidth}
      \centering %
      \hspace{1em} %
      \includegraphics[width=0.95\textwidth]{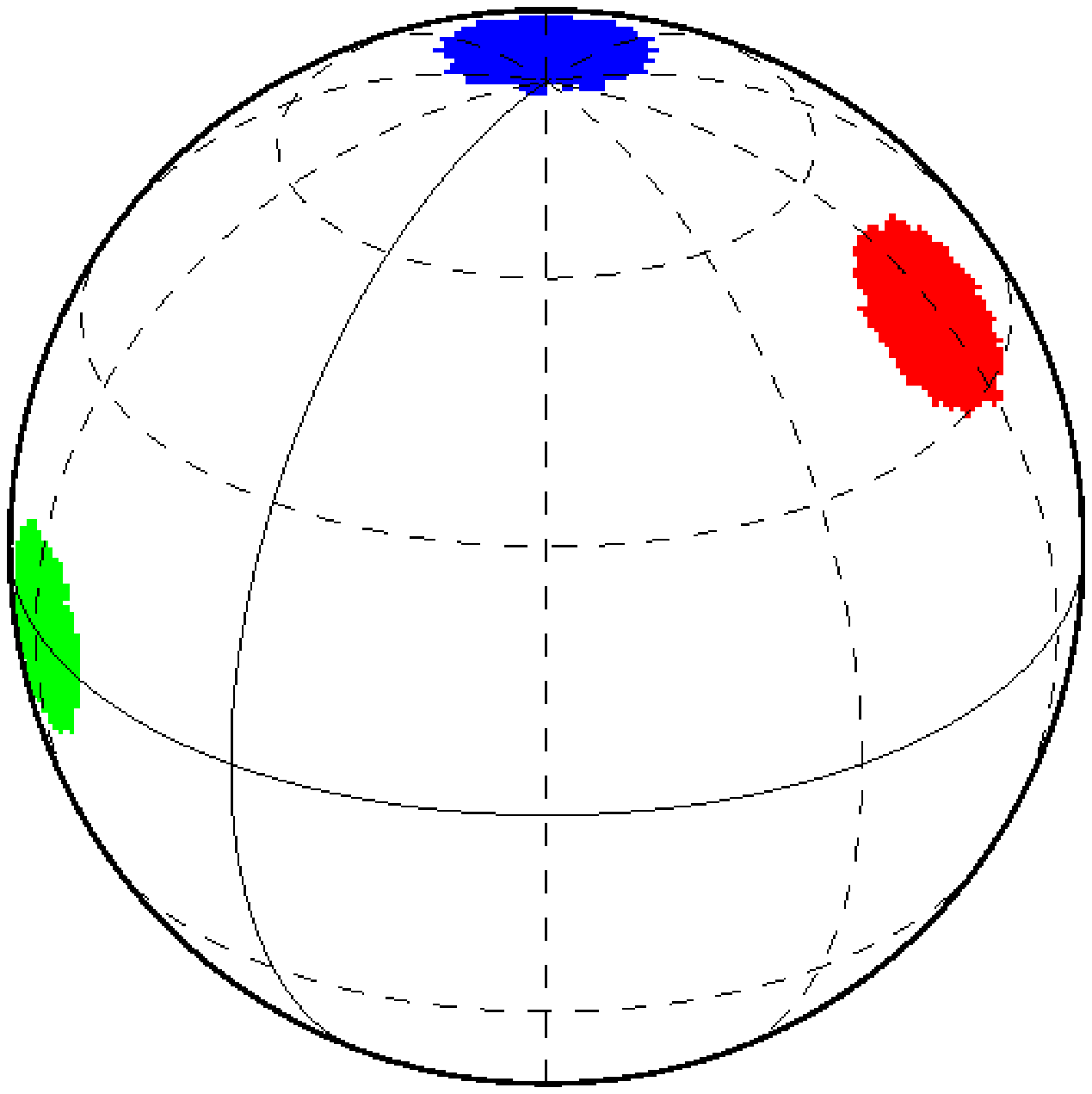}\\
      \includegraphics[width=0.95\textwidth]{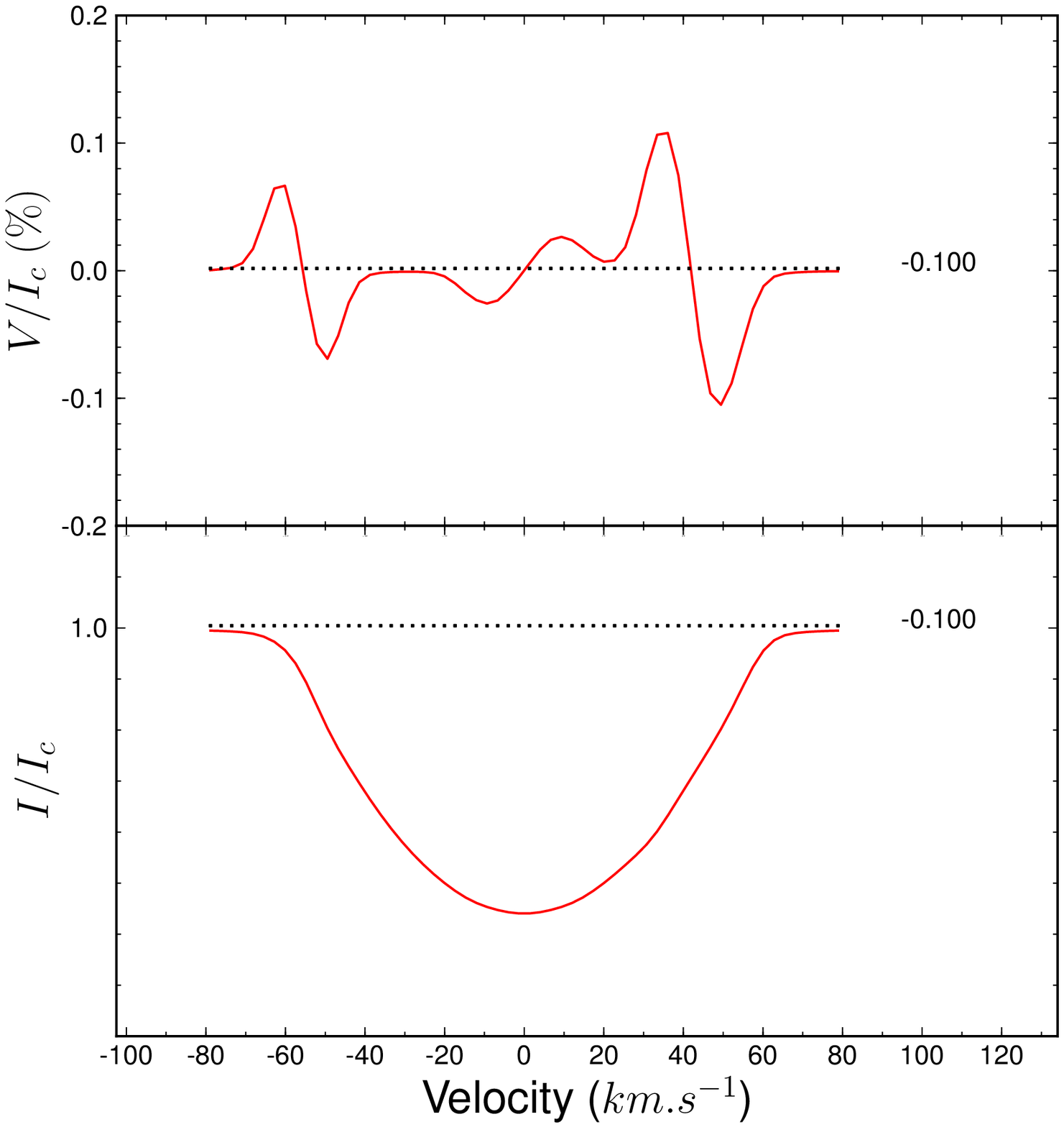}
    \end{minipage}
   } %
  \hspace{1em} %
  \subfloat[$\varphi=0.20$]{%
    \begin{minipage}{0.40\textwidth}
      \centering %
      \hspace{1em} %
      \includegraphics[width=0.95\textwidth]{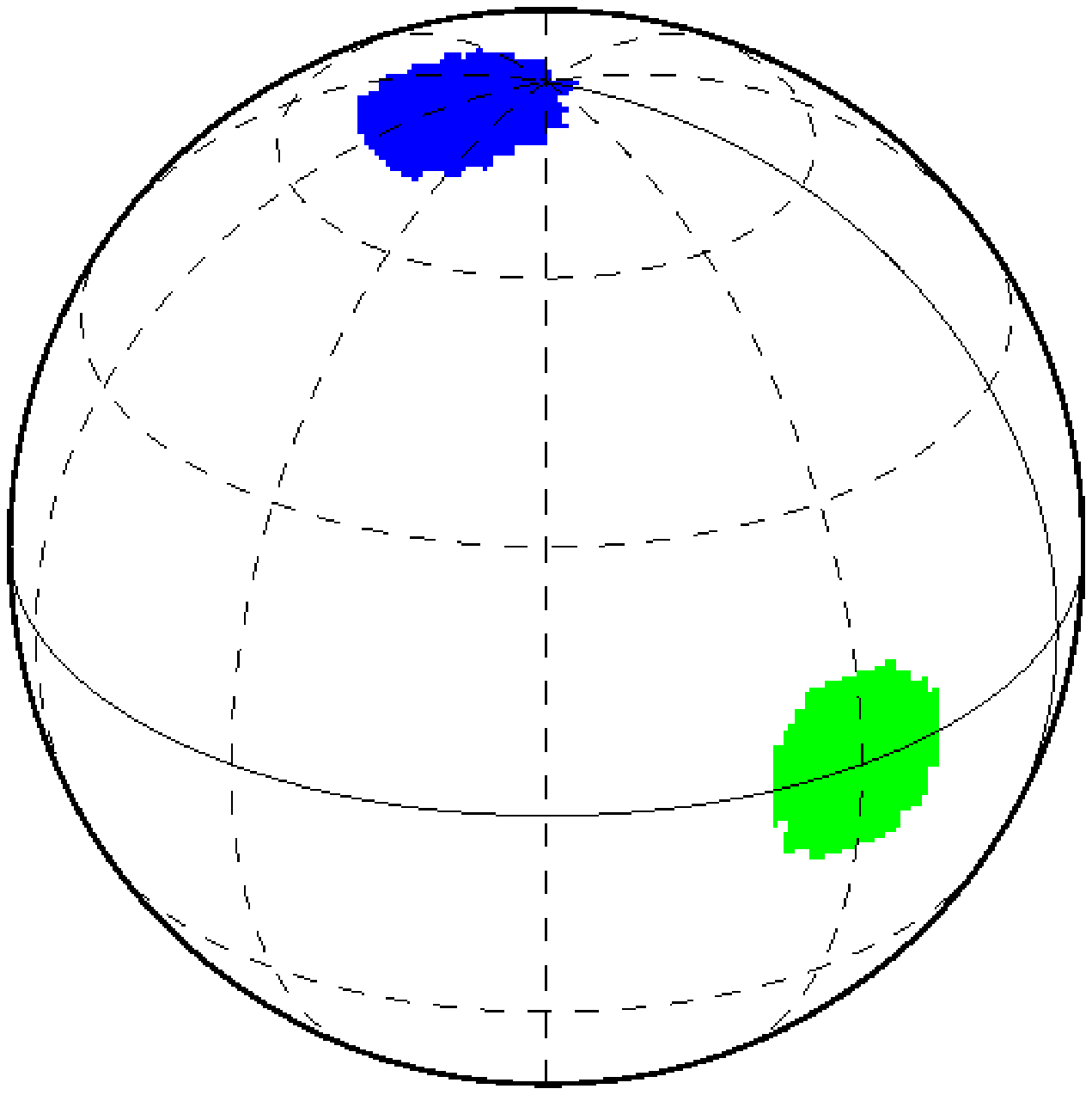}\\
      \includegraphics[width=0.95\textwidth]{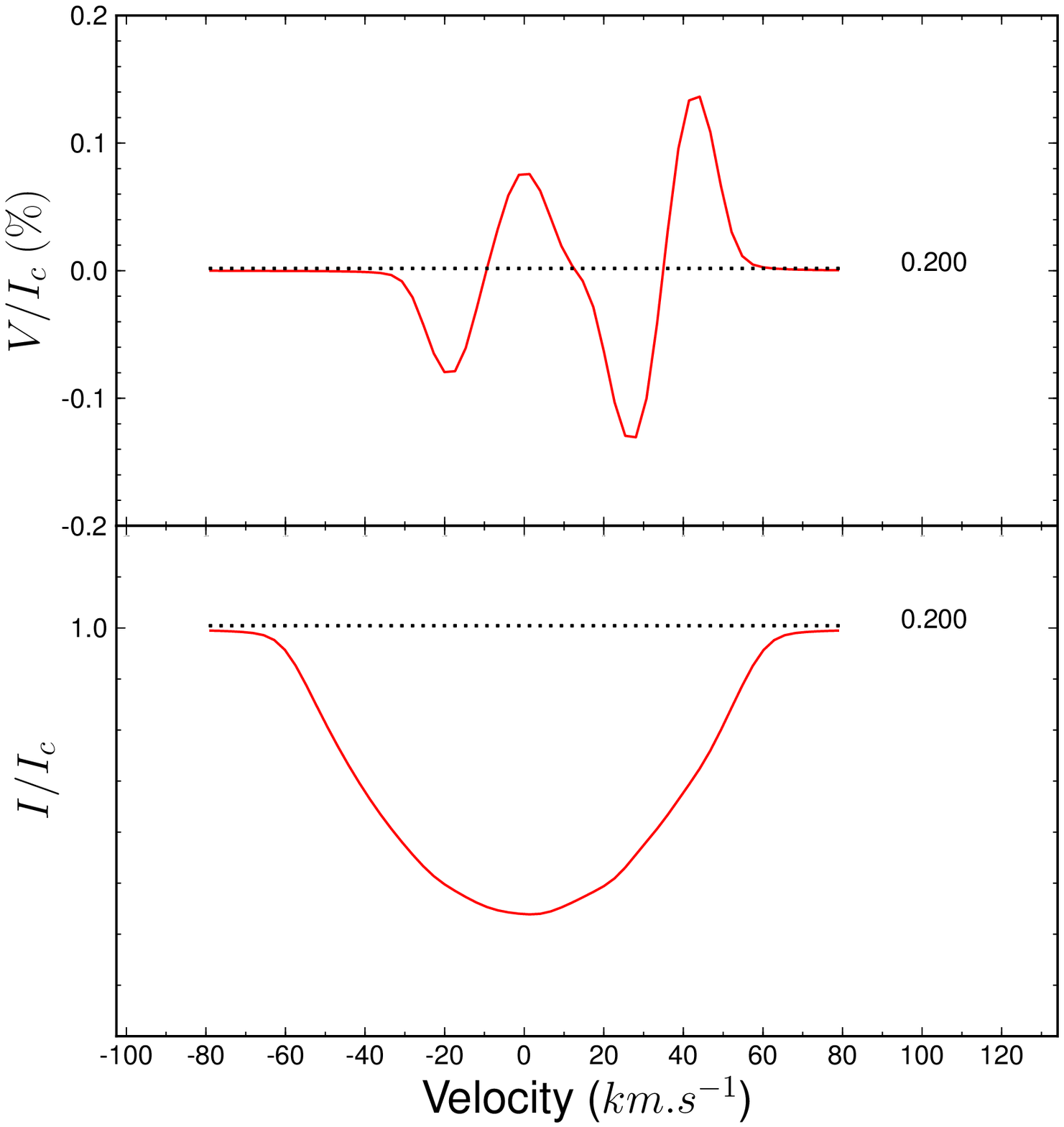}
    \end{minipage}
   }
  \end{center}
  \caption{The principles of Zeeman-Doppler Imaging (ZDI). The
favourable case of a rapidly rotating star ($\vsini=60~\kms$) for which the inclination of the
rotation axis with respect to the line of sight (LoS) is $i=60\degr$, is considered. Three
magnetic spots with $B=4~\kG$ cover each 1~\% of the stellar surface. From left to right in panel
(a) the orientation of $\vec{B}$ is azimuthal prograde (green), radial inward (blue) and radial
outward
(red). The Stokes I and V spectra are computed for a typical magnetically sensitive line using the
Unno-Rachkovsky solution of the polarised radiative transfer equations taking into account radial
velocity shifts and limb darkening, and convolved with an instrumental profile corresponding to
$R=60,000$. The star and the spectra are shown at two different epochs separated by 0.3 stellar
rotation. The main effects on which ZDI relies are well visible: \enumi\ the 3 magnetic spots are
well separated in radial velocity space, and hence they produce well-separated signatures in the
Stokes~V line profile. \enumii\ Due to different angles of $\vec{B}$ with the LoS and limb angles,
the 3 spots result in signatures with different Stokes~V amplitudes although $B=4~\kG$ for all of
them; and for a given spot the amplitude depends on the rotational phase. \enumiii\ The
high-latitude spot is always visible and always contribute to the center of the line (low values of
radial velocity), whereas low-latitude spots are visible during a fraction of the stellar rotational
cycle and the corresponding Stokes~V signatures cross most of the line profile width. \enumiv\ The
Stokes~V signatures corresponding to the radial field spots keep a constant sign during stellar
rotation, whereas for azimuthal field the signature reverses since the sign of the LoS
projection of $\vec{B}$ changes sign.}
  \label{fig:tech-zdi-principles}
\end{figure}
Spectropolarimetry consists in measuring at least one of the Stokes parameters Q, U, V as a
function of wavelength, in addition to the unpolarized intensity spectrum I.
Stokes~V signatures in spectral lines have typical amplitudes of a few $10^{-3}$ of the level
of the unpolarized continuum for active cool stars, Stokes Q and U are typically an order of
magnitude smaller\footnote{In the weak-field approximation (Eq.~\ref{eq:zeeman-weakfield}),
circular polarization (Stokes~V) is a first order effect, whereas linear polarization (Stokes Q
and U) is a second-order effect.} \cite[see ][for full Stokes polarimetry of cool
stars]{Kochukhov11}. Hence spectropolarimetric measurement require a very high
polarimetric accuracy (in particular the acquisition and reduction procedures must ensure that
instrumental polarisation is efficiently reduced, see \eg \citealt{Donati97b}), and spectra with
high signal to noise ratio (SNR). This difficulty can be overcome to some extent with multi-line
techniques which extract the polarimetric information from a large number of lines available in
stellar spectra in order to compute an average profile with increased SNR. The achieved SNR
multiplex gain can be as high as several tens when thousands of lines are used. Although some
information is lost in the averaging process,  this technique has the advantage of minimizing the
effect of blends which can be considered as a random noise. Least squares deconvolution (hereafter
LSD) is a widely used multi-line technique similar to the cross-correlation techniques used for
accurate radial velocity measurements \cite[][]{Donati97b}. LSD Stokes~V profiles can be analyzed as
a real spectral line with a reasonable accuracy for field strengths below 5~\kG\
\cite[][]{Kochukhov10} such as those encountered at the surface of low-mass stars. Alternative
multi-line methods that can be superior to LSD in some respects are being developed
\cite[\eg][]{MartinezGonzalez08, Sennhauser10}. 

The most easily derived magnetic quantity from spectropolarimetric data is the longitudinal
field $B_\ell$ --- \ie\ the line-of-sight component of the magnetic field integrated over the
visible stellar disc --- which is related to Stokes I and V through:
\begin{equation}
 B_\ell({\rm G}) = -2.14 \times 10^{11} \frac{\displaystyle\int v\,V(v)
\,{\rm d}v}{\lambda_0\,g_{\rm eff}\,c \displaystyle\int
\left[I_c-I(v)\right] {\rm d}v } \, \text{,}
\label{eq:tech-bl}
\end{equation}
where $I_c$ is the unpolarised continuum, $v$ is the radial velocity in the stellar rest frame, $c$
is the speed of light in the same unit as $v$, $\lambda_0$ is the central wavelength of the line in
nm, and $g_{\rm eff}$ is the effective Landé factor of the line \cite[][]{Rees79}.
However $B_{\ell}$ is an integral quantity which only reflects a limited fraction of the information
contained in Stokes~V line profiles.
In order to take full advantage of high-resolution spectropolarimetric observations, these are
often analyzed by means of Zeeman-Doppler
Imaging (ZDI), a method introduced by \cite{Semel89} (and analogous to Doppler Imaging,
see \eg\citealt{Vogt87}) and continuously developed over the years
\cite[\eg][]{Piskunov02, Donati06}. 
ZDI relies on three effects which result in a strong relationship between the distribution of
magnetic fields at the surface of a star and the temporal evolution of polarised signatures in
spectral lines, as shown on Fig.~\ref{fig:tech-zdi-principles}. \enumi~Due to the Doppler effect
induced by stellar rotation magnetic regions
located at different projected distances from the rotation axis contribute to different parts of a
rotationally broadened spectral line. \enumii~Different parts of the star are visible under varying
limb angles as the stars rotates. \enumiii~Zeeman-induced
polarization in spectral lines is sensitive to the orientation of the magnetic field vector (see
Sec.~\ref{sec:measurements-zeeman}). ZDI is an inverse
problem which is solved iteratively by comparing the times-series of polarized spectra computed from
a model magnetic map with the observed one until convergence is reached. As there is no unique
solution to this problem, a regularization scheme has to be used. The maximum entropy solution
corresponding to lowest magnetic energy content is often used.
ZDI studies of cool
stars rely on Stokes~V alone, Stokes Q and U are not observed due to their lower amplitudes.
\cite{Donati97a} have shown in the case of a dipolar field that a reliable ZDI reconstruction could
be performed from Stokes~V alone.

\subsection{Comparison of the two approaches}
\begin{figure}
  \begin{center}
  \leavevmode  
  \subfloat[]{%
    \begin{minipage}{0.3\textwidth}
      \centering %
      \hspace{1em} %
      \includegraphics[width=0.95\textwidth]{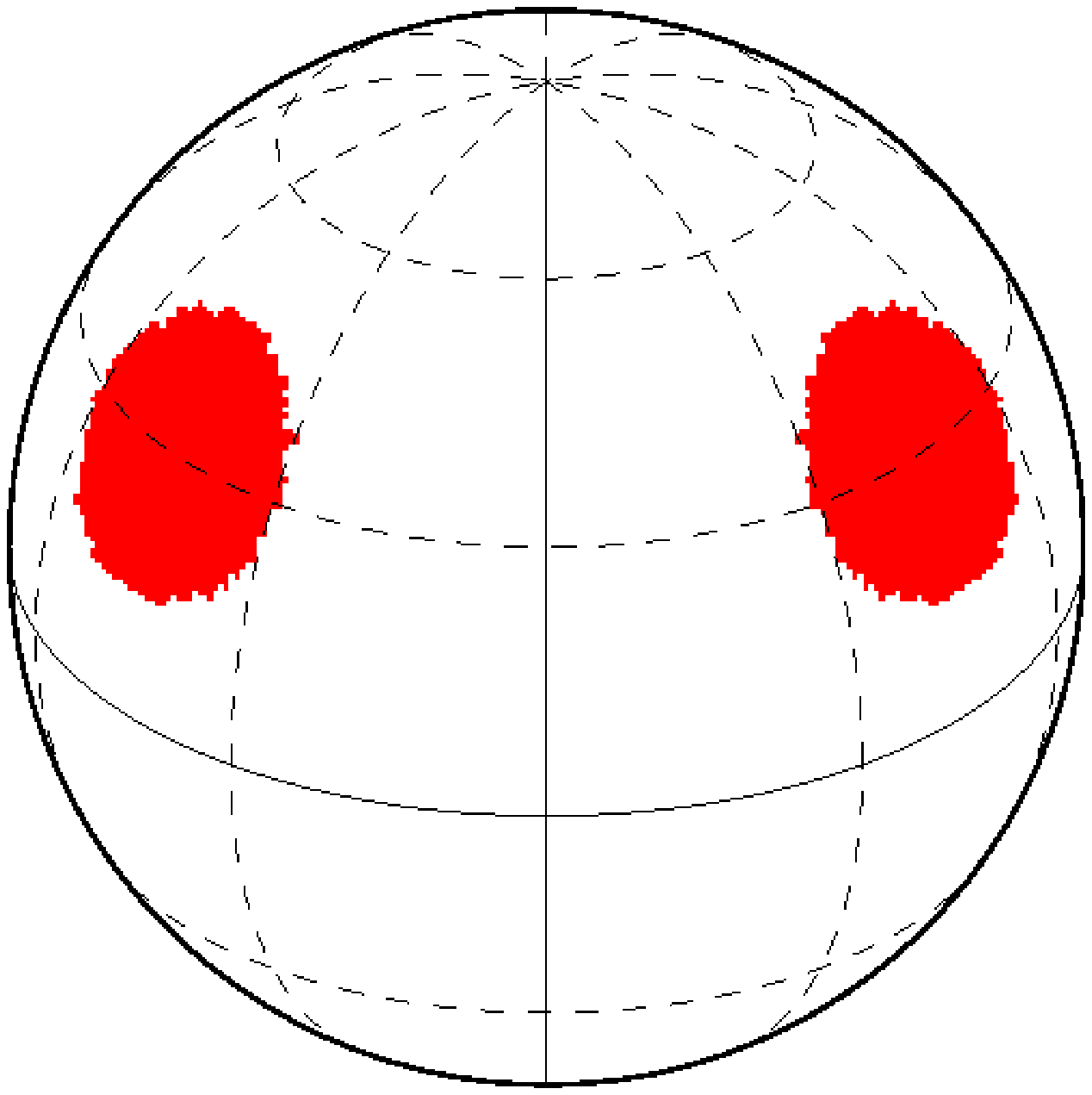}\\
      \includegraphics[width=0.95\textwidth]{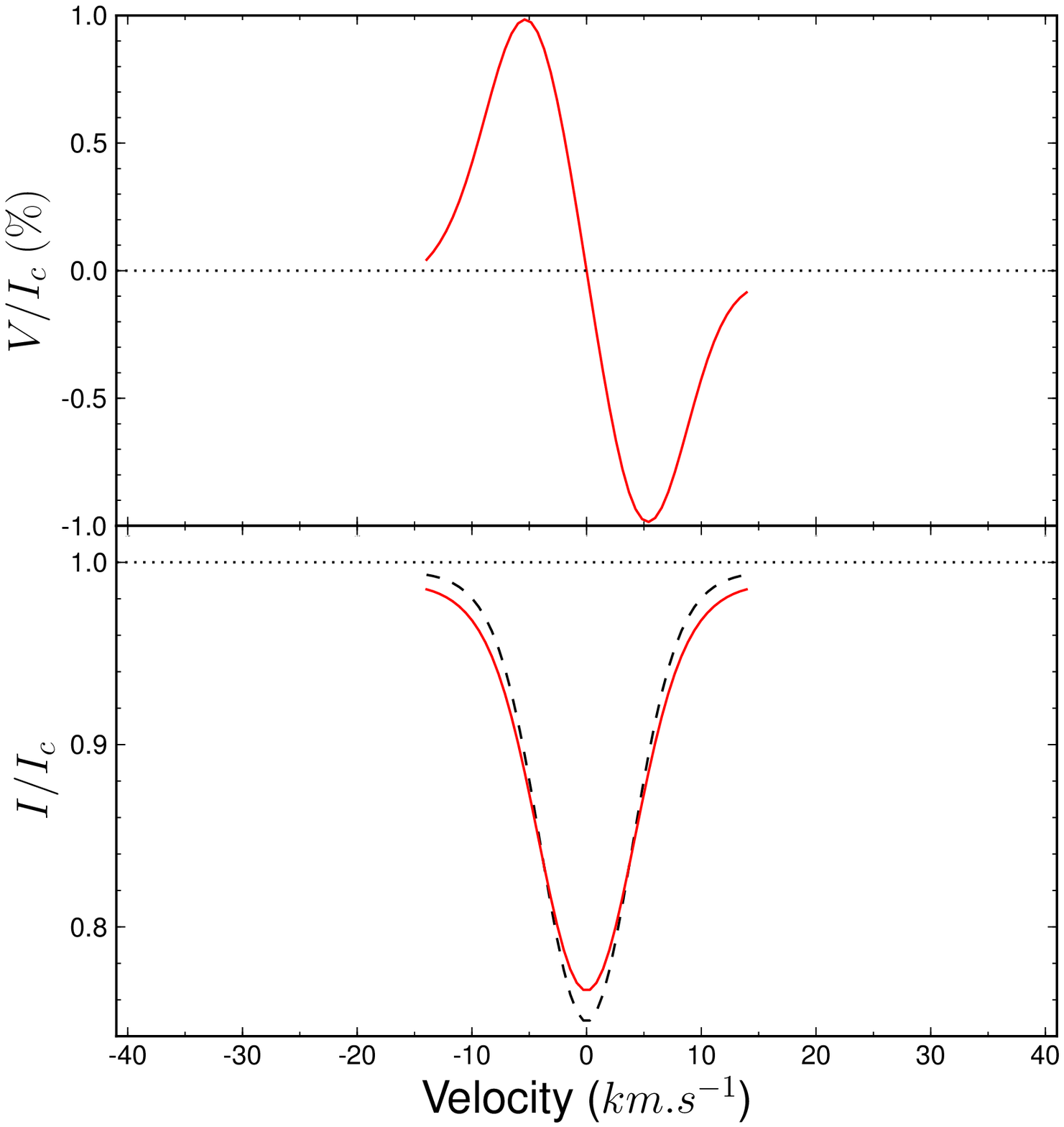}
    \end{minipage}
   } %
  \hspace{0.5em} %
  \subfloat[]{%
    \begin{minipage}{0.3\textwidth}
      \centering %
      \hspace{1em} %
      \includegraphics[width=0.95\textwidth]{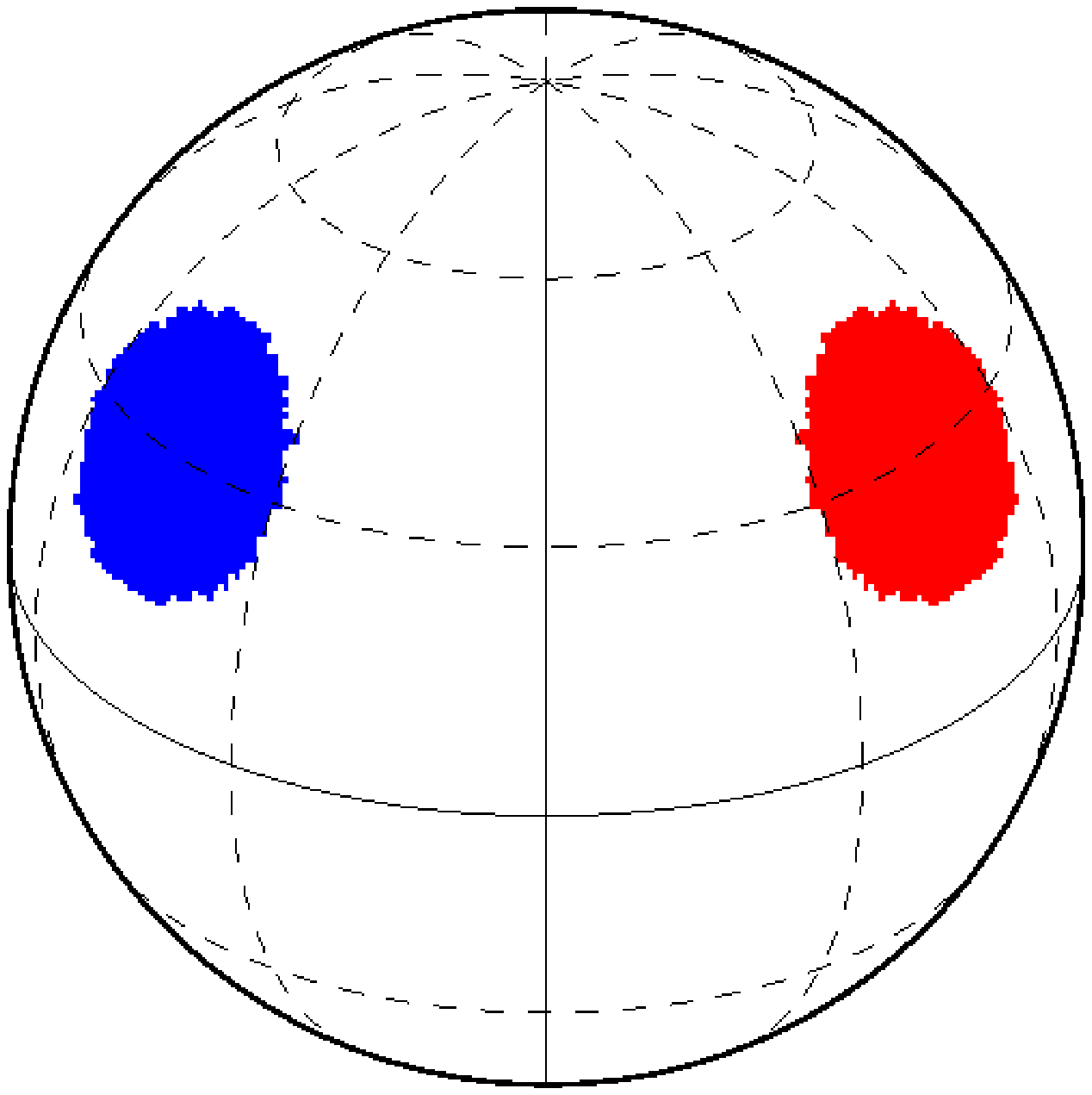}\\
      \includegraphics[width=0.95\textwidth]{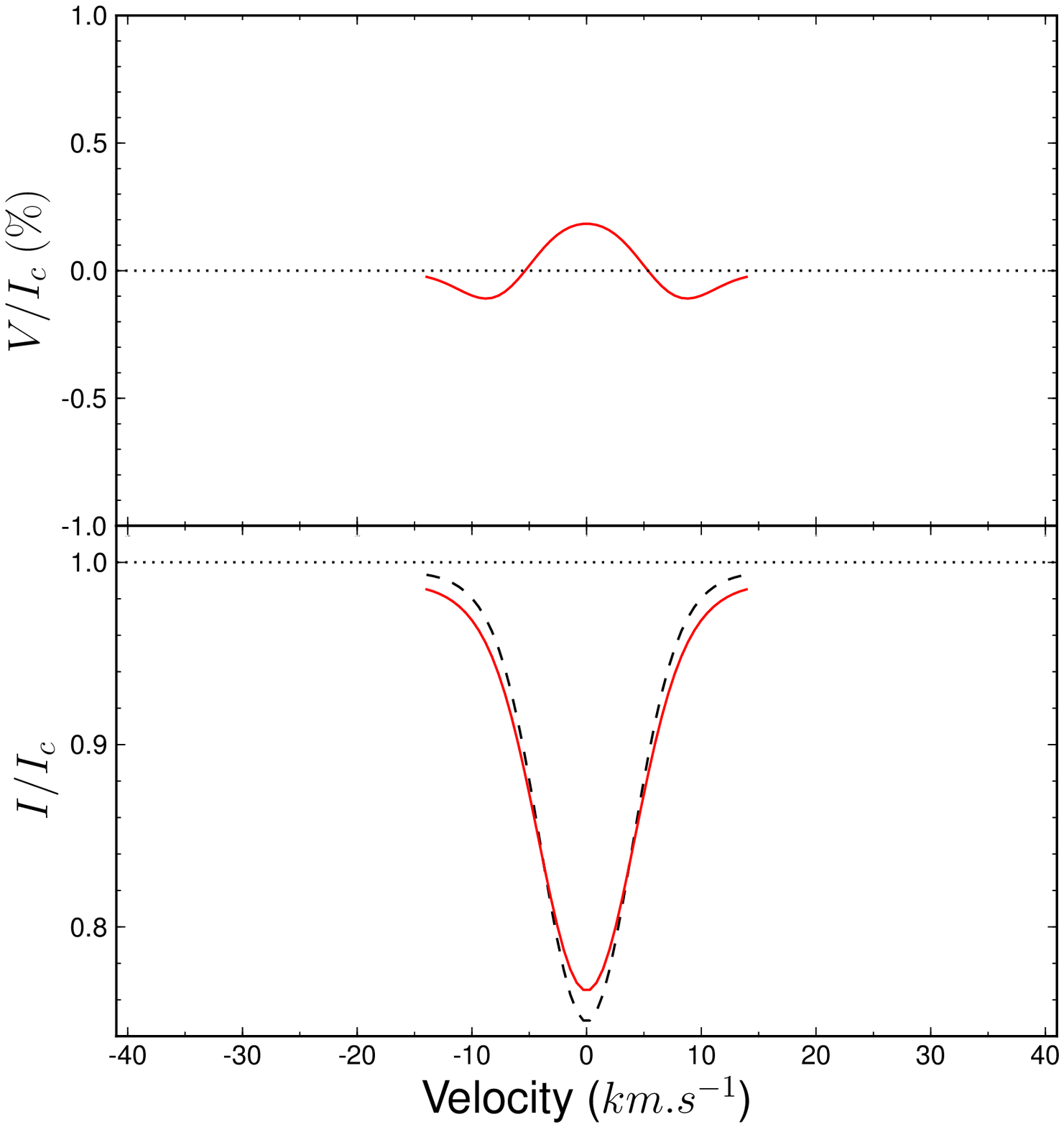}
    \end{minipage}
   } %
  \hspace{0.5em} %
  \subfloat[]{%
    \begin{minipage}{0.3\textwidth}
      \centering %
      \hspace{1em} %
      \includegraphics[width=0.95\textwidth]{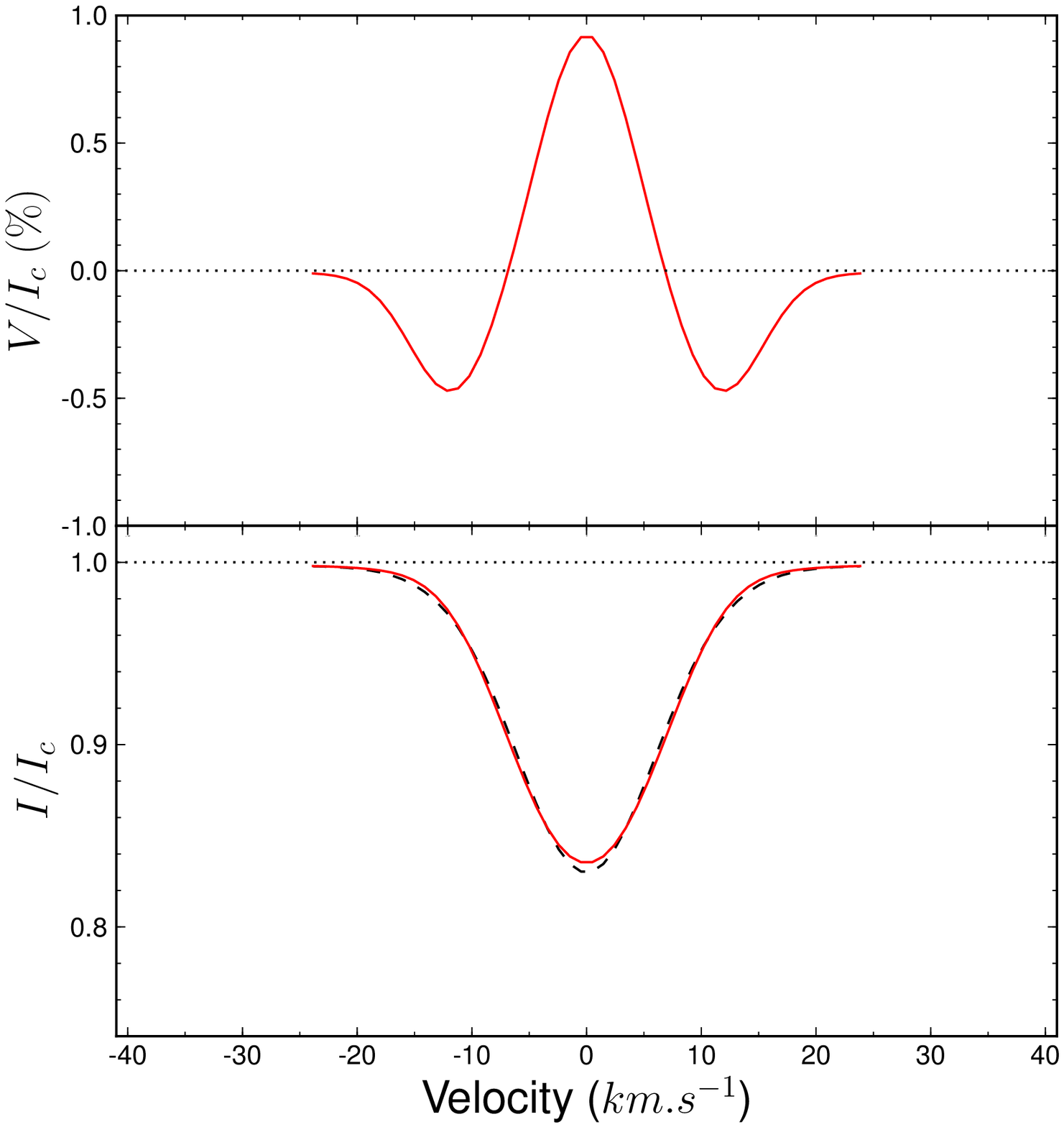}\\
      \includegraphics[width=0.95\textwidth]{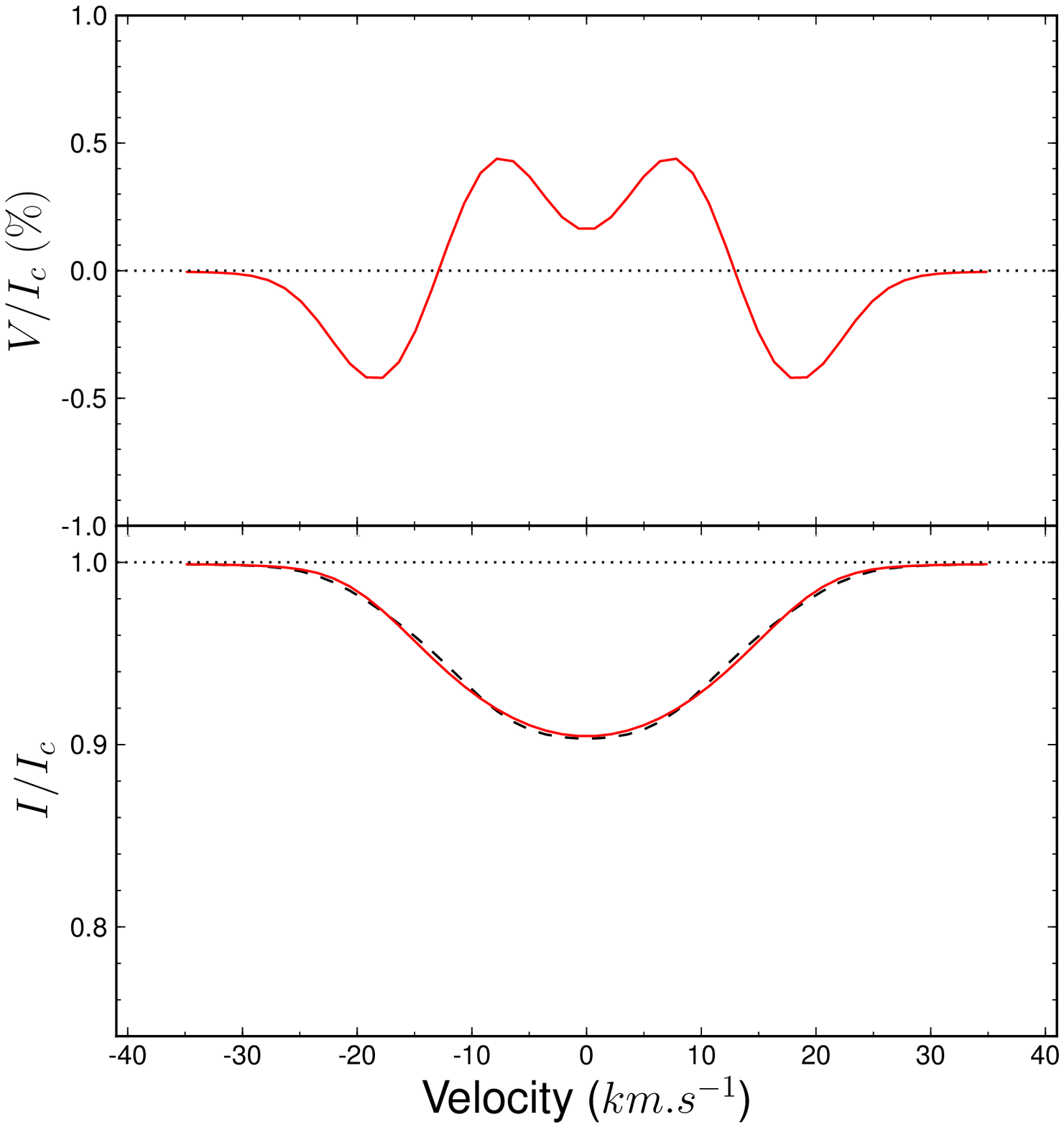}
    \end{minipage}
   }
  \end{center}
  \caption{Stokes I and V spectra are computed as in Fig.~\ref{fig:tech-zdi-principles} for two
basic magnetic field distributions. In each Stokes~I panel the black dashed line represents the
corresponding spectral line without magnetic field. \textit{(a)} Two magnetic spots with $B=4~\kG$
cover each 1~\% of the stellar surface, the field lines are radial and directed outward.
\textit{(b)} Same as \textit{(a)} but the two spots have opposite polarities: in the left spot
(blue) $\vec{B}$ is radial inward. In panels \textit{(a)} and  \textit{(b)}, $\vsini=1~\kms$. The
two magnetic field distributions cannot be distinguished from Stokes~I measurements and are thus
equally well detected. The two configurations are clearly distinguished from Stokes~V measurements,
although in case \textit{(b)} the low amplitude of circular polarisation makes it more difficult to
detect. \textit{(c)} Same as \textit{(b)} except for $\vsini=10~\kms$ (top) and $20~\kms$ (bottom).
For higher values of \vsini\ the contributions of the two spots are more easily separated in
Stokes~V spectra, whereas the Stokes~I line broadening becomes harder to detect.}
  \label{fig:tech-comparison}
\end{figure}
From measurements of the Zeeman broadening of spectral lines in unpolarized light the disc-averaged
magnetic field modulus can be estimated irrespective of the field complexity. The accuracy of the
technique is limited by the fact that the Zeeman effect has to be disentangled from other sources
of broadening and that the reference null-field profile with respect to which broadening is
estimated is only known with a limited accuracy. With spectropolarimetry  an information  on the
vector properties of the magnetic field is available, but tangled small-scale fields remain
undetected. As opposed to Zeeman broadening measurements, the null-field profile is perfectly known
-- null circular and linear polarization in spectral lines. Very weak longitudinal fields below 1~G
can be measured from LSD Stokes~V spectra provided the SNR is high enough \cite[\eg][]{Auriere09}.
Measurements of Zeeman broadening in unpolarised light and spectropolarimetry provide
complementary information on stellar magnetic fields. However it is not always
possible to obtain both measurements for a given object: spectropolarimetric measurements are
optimally performed for \vsini\ values of a few tens of \kms\ --- which offer a good compromise
between disentangling of magnetic regions of opposite polarities and line depth --- whereas line
broadening measurements are limited to approximately $\vsini<20~\kms$. In addition, both methods
are affected by temperature inhomogeneities which can ``hide'' magnetic flux.

\section{The fully convective transition}
\subsection{Magnetic fields of partly convective stars}
\label{sec:fcl-pcs}
Activity phenomena are observed on most cool stars. They are found to be variable or
even to exhibit solar-like cycles in a number of cases (see chapter by X.~Bonfils,
this book). Surface magnetic fields are directly detected for a number of partly convective stars.
Generation of magnetic fields in these partly-convective Sun-like stars is
believed to rely on similar processes as in the Sun, with an important role of
differential rotation ($\alpha\Omega$-type dynamo) and of the tachocline. In
spite of the wide variety of stellar parameters ($0.35<\mstar<1.3~\msun$
considering only main sequence objects, rotation periods and activity levels
spanning several orders of magnitude) these stars have a similar internal
structure with an inner radiative zone, a convective envelope, and likely a
tachocline in between. In addition, activity levels (expressed as luminosity in \eg X-rays or in the
\Ha\ line normalized to the stellar bolometric luminosity) of cool stars 
appear to be affected in the same way by rotation when it is expressed through
the Rossby number Ro which measures the ratio of inertia to the Coriolis
force or equivalently the ratio of the dynamical to rotational timescales
\footnote{The Rossby number is defined as: ${\rm Ro}=\frac{u_0}{\Omega
L}$ or Ro$=\frac{\Prot}{\tau_c}$, where $u_0$ is the typical fluid velocity,
$\Omega$ the rotation rate, $L$ a characteristic lengthscale (\eg the pressure
scale height), \Prot\ the rotation period, and $\tau_c$ the convective turnover
time. It is a local quantity that is expected to strongly depend on the
radial coordinate inside the star. In observational stellar
physics a global or empirical Rossby number is often used, in order to assign a
unique number to a star.} \cite[\eg][]{Mangeney84,Noyes84,Pizzolato03}.

Studies of partly-convective stars in unpolarised light have reported average magnetic fields from
the detection threshold (a few hundred Gauss typically) up to several kG \cite[\eg][]{Saar96,
Anderson10}. From these
measurements, \cite{Saar01} have found a good correlation between Rossby number and magnetic flux
($\Bf\propto Ro^{-1.2}$). Studies based on 
spectropolarimetric observations and Zeeman-Doppler Imaging have concluded that the
larges-scale component of the field is dominated by the dipole and quadrupole
modes for slow rotators, as is observed on the Sun. For faster rotators
higher order multipoles are predominant and a strong toroidal component is
detected at the photospheric level \cite[\eg][]{Petit08} --- whereas on the Sun strong toroidal
fields are thought to reside deep in the interior, mostly in the tachocline. A few
polarity reversals or even full magnetic cycles have also been observed with
these techniques, suggesting that fast rotators undergo faster magnetic cycles, although the
dependence on stellar mass is not yet clear \cite[\eg][]{Fares09, Morgenthaler11}. 

\subsection{Dynamo models of fully convective stars}
Main sequence stars with masses below $\sim 0.35~\msun$ \cite[\eg][]{Chabrier97} or
young solar-type stars (T Tauri stars, see \eg \citealt{Siess00}) are fully
convective (see Fig.~\ref{fig:convection-hrd}) and therefore do not possess a tachocline. If this
region is really a key element of the solar dynamo, fully convective stars should generate their
magnetic fields by non-solar processes.  This 
led theoreticians to model magnetic field generation in fully convective stars
without an $\Omega$-effect. \cite{Durney93} first proposed that in these
objects magnetic fields could be generated by a small-scale dynamo \ie\ 
producing magnetic field at the scale of the fluid motions. \cite{Kuker99} and
\cite{Chabrier06} performed mean-field modeling of brown dwarfs and fully
convective stars, the $\alpha^2$ dynamo resulted in steady (\ie\
showing no reversal) non-axisymmetric field dominated by high order multipoles.
Dynamo action in fully-convective stars has also been investigated with direct numerical
simulations by \cite{Dobler06} and \cite{Browning08}. In these studies,
such objects can generate strong and steady magnetic fields possessing a significant axisymmetric
component. In addition \cite{Browning08} shows that in his simulations Maxwell stresses result in 
a very low level of differential rotation.

\subsection{Magnetic fields of fully convective stars in unpolarised spectroscopy}
\label{sec:fcl-unpol}
\begin{figure}
  \begin{center}
  \leavevmode  
  \subfloat[\cite{Reiners10c}]{%
    \begin{minipage}{0.49\textwidth}
      \centering %
      \includegraphics[width=0.98\textwidth]{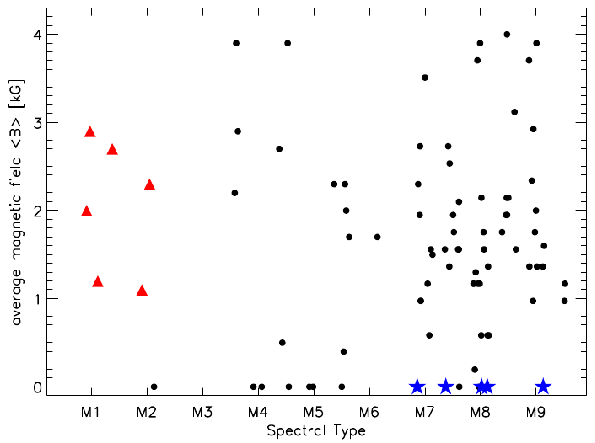}
      \label{fig:fcl-unpol-SpT-Bf}
    \end{minipage}
   } %
  \subfloat[\cite{Reiners12b}]{%
    \begin{minipage}{0.49\textwidth}
      \centering %
      \includegraphics[width=0.98\textwidth]{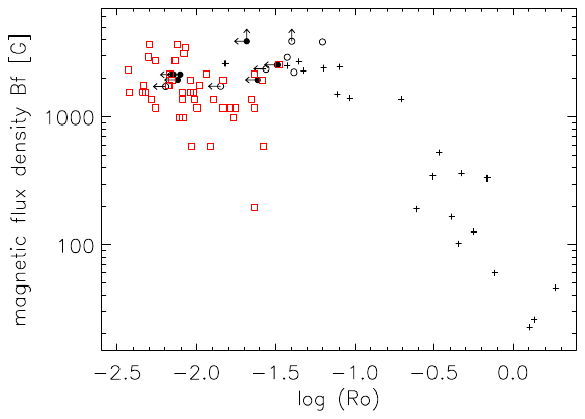}
      \label{fig:fcl-unpol-Ro-Bf}
    \end{minipage}
   }
  \end{center}
  \caption{\textit{(a)} Average magnetic fields of M dwarfs measured from Stokes~I as a function of
  spectral type. Young early M dwarfs are depicted as red triangles, field M dwarfs as black
  circles and young accreting brown dwarfs as blue stars.\textit{(b)} Average magnetic fields of
  G--K--M dwarfs derived from Stokes~I measurements as a function of Rossby number.  Crosses are
  Sun-like stars  \cite[][]{Saar96,Saar01}, circles are M-type of spectral class M6 and earlier
  \cite[][]{Reiners09a}. For the latter, no period measurements are available and Rossby numbers are
  upper limits (they may shift to the left hand side in the figure). The black crosses and circles
  follow the rotation-activity relation known from activity indicators. Red squares are objects of
  spectral type M7--M9 \cite[][]{Reiners10a} that do not seem to follow this
  trend ($\tau_c$ = 70~\d
  was assumed for this sample).
 }
  \label{fig:fcl-unpol}
\end{figure}

Magnetic fields of main sequence M dwarfs have been investigated since the 1980s in unpolarised
spectroscopy, using atomic lines \cite[][]{Saar85,Saar94,Johns-Krull96, Kochukhov09}. However, most
\Bf\ values now available have been obtained since \cite{Reiners06b} have proposed a method to
perform such measurements based on spectral lines of the FeH molecule (see
Sec.~\ref{sec:tech-unpol}). Measurements performed on stars located close to the fully convective
limit are presented in \cite{Reiners07,Reiners09b,Reiners09a}. Measured $\Bf$ values are plotted as
a function of spectral type on Fig.~\ref{fig:fcl-unpol-SpT-Bf}, they span the range 0--4~\kG. No
strong break can detected at the spectral type at which stars become fully convective (M3--M4),
similarly to what is observed for activity proxies (in particular for X-ray and \Ha\ emission, see
chapter by X.~Bonfils, this book). \enumi\ The range of magnetic fluxes measured is similar to more
massive solar-type stars. \enumii\ The observed scatter can be accounted for by the effect of
rotation, there is evidence for a dependence on Rossby number similar to that observed for
solar-type stars for stars of spectral type M0--M6 \cite[see][and
Fig.~\ref{fig:fcl-unpol-Ro-Bf}]{Reiners09a}: \Bf\ increases towards low Ro, and then reaches a
saturation level ($\Bf\sim2-4~\kG$) for Ro$\sim 0.1$ (\ie\ $\Prot\sim6~\d$ for a M3 dwarf).
\subsection{Magnetic fields of fully convective stars in spectropolarimetry}
  \label{sec:fcl-pol}
Following the first detection in polarised light of a large-scale magnetic field on a fully
convective star by \cite{Donati06}, a spectropolarimetric survey of a small sample of active
M~dwarfs lying on both sides of the fully convective boundary has been carried out.
\begin{figure}
\begin{center}
  \includegraphics[angle=270,width=0.75\textwidth]{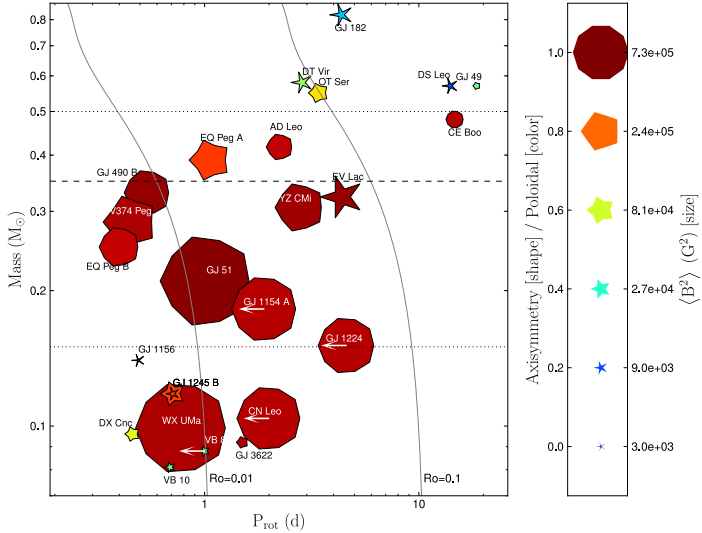}%
  \caption{Properties of the large-scale magnetic fields of the sample of M dwarfs observed in
  spectropolarimetry as a function of
  rotation period and mass. Larger symbols indicate stronger fields, symbol
  shapes depict the degree of axisymmetry of the reconstructed magnetic field
  (from decagons for purely axisymmetric to sharp stars for purely non
  axisymmetric), and colours the field configuration (from blue for purely
  toroidal to red for purely poloidal). Solid lines represent contours
  of constant Rossby number $Ro=0.1$ (saturation threshold) and $0.01$. The
  theoretical full-convection limit ($\mstar \sim0.35~\msun$) is plotted as a
  horizontal dashed line, and the approximate limits of the three stellar
  groups discussed in the text are represented as horizontal solid lines. White arrows mark stars
  for which only an upper limit for the rotation period is known. In these cases the field geometry
  is assumed, and the magnetic energy extrapolated from the longitudinal field values. Adapted from
  \cite{Morin10a}.}
  \label{fig:fcl-plotMP}
\end{center}
\end{figure}
The main results of this study are presented on Fig.~\ref{fig:fcl-plotMP}, showing the main
properties of the large-scale magnetic fields reconstructed with ZDI as a function of stellar
mass and rotation period. A sharp change is observed close to the boundary to full-convection
\cite[][]{Morin08a, Morin08b, Donati08b, Phan-Bao09}:

\begin{itemize}
\item M dwarfs more massive than $\sim0.5~\msun$ (partly convective) exhibit
large-scale magnetic fields with a strong toroidal component, even dominant in some
cases; the poloidal component is strongly non-axisymmetric. For most of these
stars, surface differential rotation can be derived, values are comprised between once and twice the
solar rate approximately, and the magnetic fields evolve beyond recognition on a timescale of a few
months. These properties are reminiscent of the observations of more massive (G and K) active
stars (\cf\ Sec.~\ref{sec:fcl-pcs}).
\item Stars with masses between $\sim$0.2 and 0.5~\msun\ (close the fully
convective limit) host much stronger large-scale magnetic field with radically
different geometries: almost purely poloidal, generally nearly axisymmetric,
always close to a dipole more or less tilted with respect to the rotation axis.
These magnetic field distributions are observed to be stable on timescales of several years, and
differential rotation (when measurable) is of the order or a tenth of the solar
rate.  These findings are in partial agreement with the recent numerical study by
\cite{Browning08}. Similarly, it is observed that fully convective stars can
generate strong and long-lived large-scale magnetic fields featuring a strong
axisymmetric component, that are able to quench differential rotation. But
almost purely poloidal surface magnetic fields are reported, whereas in the
simulation the axisymmetric component of the field is mainly toroidal (although
the simulation does not encompass the stellar surface). 
\item Much stronger large-scale magnetic fields are measured from Stokes~V spectra for
fully-convective stars with respect to partly-convective ones, whereas the typical \Bf\ value
derived from unpolarised measurement does not seem to be affected by the fully-convective
transition. Hence the ratio of the magnetic fluxes recovered from Stokes~V to the one recovered from
I increases across the fully-convective boundary. This has been interpreted as evidence that
magnetic fields of fully convective stars are organized on larger spatial scales than that of
partly-convective ones.
\end{itemize}
It is worth noting that the sample does not allow to definitely disentangle the effects of mass and
rotation on stellar magnetic fields. Further observations are needed to address this issue.

\section{Very low mass stars and brown dwarfs}
\label{sec:vlms}
\subsection{Dynamo models of ultracool dwarfs}
\label{sec:vlms-act-dynamo}
Except for the energy production mechanism, late M and early L brown dwarfs are similar to
very-low-mass stars: they are fully-convective objects with a highly conducting interior and are
expected to sustain dynamo action in a similar way as fully-convective stars
do \cite[][]{Chabrier06}. However, two important differences between these ultracool dwarfs and
fully-convective mid M dwarfs exist. \enumi\ In ultracool dwarfs an outer layer with low electric
conductivity is present at the surface. \cite{Mohanty02} have shown that this layer could prevent
the generation of magnetic stresses that are required to sustain a chromosphere. A decrease of
the activity level measured in the \Ha\ line has indeed been measured, as well as in X-rays (see
chapter by X.~Bonfils, this book). However radio luminosity remains at a high level \cite{McLean12},
in agreement with the mean-field modelling by \cite{Chabrier06} who conclude that the generation of
magnetic fields through large-scale dynamo action should still be efficient in ultracool dwarfs.
\enumii\ In a significant fraction of ultracool dwarfs dynamo likely operate in a very low Rossby
number regime, \ie\ strongly dominated by rotation. Indeed rotational braking appears to be very
inefficient in these objects, with $\vsini>10~\kms$ (\ie\ $\Prot<0.5~d$) often measured
\cite[][]{Reiners10a, McLean12}; and convective turnover times are expected to be longer than in
hotter objects from models \cite[][]{Chabrier06} as well as from observations of activity
\cite[][]{Kiraga07}.

\subsection{Magnetic fields of ultracool dwarfs in unpolarised light}
\label{sec:vlms-unpol}
Zeeman-induced broadening of spectral lines in ultracool dwarfs has so far only been investigated
from molecular lines of FeH. \cite{Reiners10a} present the analysis of rotation, magnetic field and
activity of a volume-limited sample ($d<20~$pc) of M7--M9.5 dwarfs. Strong magnetic fields
(up to 4~\kG) are detected down to the lowest temperatures probed (see
Fig.~\ref{fig:fcl-unpol-SpT-Bf}). However, as opposed to M0--M6 dwarfs discussed in
Sec.~\ref{sec:fcl-unpol}, a number of stars rotating faster than 10~\kms\ ($\log{Ro}\lesssim-1.5$)
exhibit average magnetic of the order of 1~\kG\ (see Fig.~\ref{fig:fcl-unpol-Ro-Bf}). This
observation is interpreted as a breakdown of the magnetic field saturation at low Rossby numbers for
ultracool dwarfs. Among 4 young accreting brown dwarfs with similar spectral types,
\cite{Reiners09c} also found systematically weak magnetic fields (upper limits of a few hundred
Gauss), much weaker than what has been measured for young accreting low-mass stars
\cite[][]{Johns-Krull07}. These observations have not yet been explained by theoretical models.

\subsection{Magnetic fields of ultracool dwarfs in spectropolarimetry}
\label{sec:vlms-pol}
\cite{Morin10a} have studied a small sample of mid-to-late M dwarfs, including 9 with
$\msun<0.15~\msun$ (mostly at spectral types M5--M6), using spectropolarimetry. For 5 stars a ZDI
study could be performed. Two different types of magnetism are found: two stars possess a strong
large-scale magnetic field, predominantly dipolar (similar to those described in
Sec.~\ref{sec:fcl-pol}); whereas for three stars the large-scale field is much weaker and complex.
For the four remaining objects, the Stokes~V signatures and derived $B_\ell$ values indicate that
such a dichotomy also exists. This is all the more surprising that all these stars have very similar
stellar parameters ($0.08<\mstar<0.15~\msun$ and $0.4<\Prot<1.5~\d$ for stars for which a period
measurement is available), see Fig.~\ref{fig:fcl-plotMP}, bottom-left corner.  Such a behaviour had
not been foreseen, and it is not yet clear whether another parameter than mass and rotation (such as
age) plays an important role in determining the magnetic topology of stars, or if two types of
dynamo genuinely coexist in this range of parameters.  The 6 stars hosting a weak and multipolar
large-scale field have been studied by \cite{Reiners07} and \cite{Reiners09a} who derived
$\Bf>1~\kG$ for all of them. Hence the precise relationship between the two types of magnetism
observed in spectropolarimetry and the fade of the rotation-dominated dynamo inferred from
unpolarised spectroscopy (Sec.~\ref{sec:vlms-unpol}) is still unclear.

\subsection{Activity--magnetic field relations in ultracool dwarfs}
\begin{figure}
  \centering%
  \includegraphics[width=0.8\textwidth]{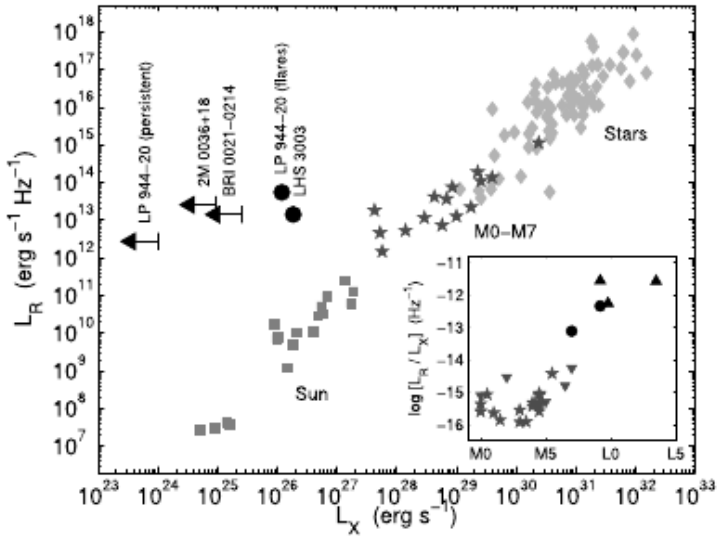} 
  \caption{Radio luminosity as a function of X-ray luminosity for cool stars and solar flares.
    Both quantities correlate very well for spectral types earlier than M7. For later spectral
    types $L_R$ remains almost constant whereas $L_X$ drops. From \cite{Berger06}.}
  \label{fig:vlms-LX-LR}
\end{figure}

The various activity proxies appear to correlate reasonably well with each other, as well as with
rotation and magnetic field strength for most cool stars (see lecture notes by X.~Bonfils, this
book). However, these correlations appear to partly break-down at late M spectral types. As
mentioned in Sec.~\ref{sec:vlms-act-dynamo}, the growing neutrality of the outer atmosphere
towards low temperatures seems to inhibit chromospheric emission \cite[\eg][]{Mohanty03}.
\cite{Reiners10a} have shown that $L_{\rm H\!\alpha}/L_{bol}$ still correlates with the surface
magnetic flux for ultracool dwarfs, although the correlation strongly depends on spectral type. As
far as coronal emission is concerned, a remarkable correlation between X-ray and radio luminosities
(respectively
associated with thermal and non-thermal electron populations) exists for cool stars. This
Güdel--Benz relation is valid over several order of magnitudes for quiescent and flaring integrated
stellar luminosities as well as for solar flares \cite[][]{Guedel93, Benz94}. However it breaks down
at spectral types later than M7: the observed X-ray luminosities (or upper limits) drop whereas
radio luminosities remain at a roughly constant level for spectral types as late as mid L (see
Fig.~\ref{fig:vlms-LX-LR}, and \citealt{Berger06}) and hence ${L_R}/{L_{bol}}$ steeply increases
towards late spectral types.
The observations of polarised radio pulses
at frequencies several GHz on a mid M dwarf and ultracool dwarfs have been interpreted as electron
cyclotron maser instability emission \cite[][]{Hallinan08}, similar to what is detected on giant
planets of the solar system \cite[see \eg][]{Zarka98}. The observations are consistent with the
present presence of a strong dipolar component of the magnetic field (polar fields of the order of
one to a few \kG) slightly tilted with respect to the rotation axis, in good agreement with
spectropolarimetric observations \cite[][]{Hallinan09,McLean11}. 

\section{Concluding remarks}
\subsection{A dynamo continuum low-mass stars, brown dwarfs, planets ?}
The large-scale magnetic fields of a number of M dwarfs hosting a strong axial dipole
component (see Sec.~\ref{sec:fcl-pol}) are more reminiscent of the magnetic field of Jupiter rather
than that of active solar-type stars. The parallel is further supported by theoretical studies:
\cite{Goudard08} have shown that a planetary dynamo simulation could switch from an Earth-like
dipolar field to propagating dynamo waves reminiscent of the solar magnetic field simply by
changing the aspect ratio of the convective region.
Additionally, \cite{Christensen09} showed that the scaling law between surface magnetic field
strength and convective energy density initially derived by \cite{Christensen06} from geodynamo
simulations, can account for the observed field strengths of a number of objects including the
Earth, Jupiter as well as rapidly rotating main sequence M dwarfs and young T Tauri stars.
The concept of a dynamo continuum from planets to low-mass stars is already at the root of several
studies \cite[\eg][]{Reiners10b,Morin11,Schrinner12} and will likely result in new advances in the
forthcoming years.

\subsection{Magnetic fields of young Suns}
The focus of these notes has been mostly on main sequence M dwarfs. Young solar type stars (ages
of a few Myr), called T Tauri stars, also undergo a fully convective phase. Strong magnetic fields
are detected on these objects both in unpolarised spectroscopy \cite[\eg][]{Johns-Krull07} and in
spectropolarimetry \cite[\eg][]{Donati08a}. It is likely that these objects generate their magnetic
fields through dynamo processes similar to those acting in main sequence M dwarfs. This idea is
further supported by the observation of a similar sequence of magnetic topologies from multipolar to
dipolar towards thicker convection zones (relative to the stellar radius), and the possible
existence of a domain where dipolar and multipolar fields coexist among T Tauri stars (Gregory,
\ea\ 2012, submitted).

\subsection{The need for combined observations}
As evidenced throughout these notes, different approaches exist to study magnetism and activity in
low-mass stars and brown dwarfs. Activity phenomena are indirect proxies,
their relationships with magnetic fields rely on a complex physics and depends on stellar
properties. Direct measurements based on the Zeeman effect allow us to directly probe the magnetic
field at photospheric level. However such measurements based on unpolarised spectroscopy and
spectropolarimetry provide us with different and complementary information that can generaly not be
compared in a straightforward way, and have their own limitations. Much remains to be understood
about the magnetism of low-mass stars, and even more for brown dwarfs. The ability to combine
information stemming from different observational approaches is likely one of the ways towards
future progress.

\Appendix
\section{Poloidal and toroidal fields}
A solenoidal (\ie\ divergence-free) vector field can be decomposed uniquely into the sum of a
poloidal and a toroidal components, according to:
\label{sec:pol-tor}
\begin{equation}
  \label{eq:pol-tor}
  \vec{B}=\vec{B_{\rm pol}}+\vec{B_{\rm tor}}%
  = \nabla\times\nabla\times(P\vec{e_r}) + \nabla\times(T\vec{e_r})
\end{equation}
In spherical coordinates, a poloidal field can have all 3 components non identically zero, whereas
a toroidal vector field lies on a sphere \ie\ it has no radial component. In the particular case of
an axisymmetric vector field,  $\vec{B}_{\rm pol, axi}$ = $(B_r, B_{\theta}, 0)$ (no azimuthal
component), whereas $\vec{B}_{\rm tor, axi}$ = $(0, 0, B_{\phi})$ (purely azimuthal). For such a
solenoidal axisymmetric vector field, the poloidal--toroidal decomposition reads:
\begin{equation}
  \label{eq:pol-tor-pot}
  \vec{B}_{\rm axi} = %
  \underbrace{\nabla\times({A\,\vec{e_{\phi})}}}_%
    {\text{poloidal part}} + %
  \underbrace{B\,\vec{e_{\phi}}}_{\text{toroidal part}}%
\end{equation}
The reader is referred to \cite{Chandra61} for a more detailed explanation.
\def\degr{\hbox{$^\circ$}}
\newcommand{\Ia}[1]{\mbox{$I_{#1\degr}$}}

\section{Stokes parameters}
\label{sec:stokes-param}
Spectropolarimetry consists in studying the polarisation of a radiation (\ie\ the temporal
evolution of the electric and magnetic field vectors in the wavefront plane) as a function a
wavelength. A complete description of an electromagnetic wave can be obtained with the four
independent Stokes parameters:
\begin{itemize}
  \item $I$ is the unpolarised light intensity 
  \item $Q$ and $U$ are the two orthogonal states which describe linear polarisation
  \item $V$ measures the net circular polarisation.
\end{itemize}
These parameters can be simply defined from an idealized measurement procedure. Let's consider an
ideal linear polarising filter and two ideal circular right- and left-handed polarising filters.
We consider a incident radiation to be characterized, and a reference direction (in the wavefront
plane) for the measurement of linear polarisation. We note $\Ia{\alpha}$ the intensity
measured through perfect linear polarizer with a polarizing  direction making an angle $\alpha$
with respect to the reference direction. $I_{\circlearrowleft}$ and $I_{\circlearrowright}$
represent the intensity measured through perfect circular polarizer selecting respectively
right-hand and left-hand circular polarisation. \Ia{0}, \Ia{45},\Ia{90}, \Ia{135}, 
$I_{\circlearrowleft}$ and $I_{\circlearrowright}$ are successively measured. The Stokes parameters
are then defined as shown in Fig.~\ref{fig:stokes-param}, see also \cite{Landi92}.

\begin{figure}[h!]
\begin{center}
  \psset{unit=0.6cm}
  \begin{pspicture}(0,0)(15,-9)
  \rput(0,-2){$I=$}
  \pscircle[linecolor=gray](1.25,-2){0.75}
  \psline[linecolor=red,linewidth=2pt]{<->}(1.25,-2.75)(1.25,-1.25)
  \rput[b](1.25,-1){$\Ia{0}$}
  \rput(2.5,-2){$+$}
  \pscircle[linecolor=gray](3.75,-2){0.75}
  \psline[linecolor=gray](3.75,-2.75)(3.75,-1.25)
  \psline[linecolor=red,linewidth=2pt]{<->}(3,-2)(4.5,-2)
  \rput[b](3.75,-1){$\Ia{90}$}
  \rput(5,-2){$=$}
  \pscircle[linecolor=gray](6.25,-2){0.75}
  \psline[linecolor=gray](6.25,-2.75)(6.25,-1.25)
  \psline[linecolor=red,linewidth=2pt]{<->}(5.72,-1.47)(6.78,-2.53)
  \rput[b](6.25,-1){$\Ia{45}$}
  \rput(7.5,-2){$+$}
  \pscircle[linecolor=gray](8.75,-2){0.75}
  \psline[linecolor=gray](8.75,-2.75)(8.75,-1.25)
  \psline[linecolor=red,linewidth=2pt]{<->}(8.22,-2.53)(9.28,-1.47)
  \rput[b](8.75,-1){$\Ia{135}$}
  \rput(10,-2){$=$}
  \psarcn[linecolor=red,linewidth=2pt]{->}(11.25,-2){0.75}{0}{360}
  \rput[b](11.25,-1){$I_{\circlearrowright}$}
  \rput(12.5,-2){$+$}
  \psarc[linecolor=red,linewidth=2pt]{->}(13.75,-2){0.75}{0}{360}
  \rput[b](13.75,-1){$I_{\circlearrowleft}$}
  \rput(0,-5){$Q=$}
  \pscircle[linecolor=gray](1.25,-5){0.75}
  \psline[linecolor=red,linewidth=2pt]{<->}(1.25,-5.75)(1.25,-4.25)
  \rput[b](1.25,-4){$\Ia{0}$}
  \rput(2.5,-5){$-$}
  \pscircle[linecolor=gray](3.75,-5){0.75}
  \psline[linecolor=gray](3.75,-5.75)(3.75,-4.25)
  \psline[linecolor=red,linewidth=2pt]{<->}(3,-5)(4.5,-5)
  \rput[b](3.75,-4){$\Ia{90}$}
  \rput(7.5,-5){$U=$}
  \pscircle[linecolor=gray](8.75,-5){0.75}
  \psline[linecolor=gray](8.75,-5.75)(8.75,-4.25)
  \psline[linecolor=red,linewidth=2pt]{<->}(8.23,-4.47)(9.28,-5.53)
  \rput[b](8.75,-4){$\Ia{45}$}
  \rput(10,-5){$-$}
  \pscircle[linecolor=gray](11.25,-5){0.75}
  \psline[linecolor=gray](11.25,-5.75)(11.25,-4.25)
  \psline[linecolor=red,linewidth=2pt]{<->}(10.73,-5.53)(11.78,-4.47)
  \rput[b](11.25,-4){$\Ia{135}$}
  \rput(0,-8){$V=$}
  \psarcn[linecolor=red,linewidth=2pt]{->}(1.25,-8){0.75}{0}{360}
  \rput[b](1.25,-7){$I_{\circlearrowright}$}
  \rput(2.5,-8){$-$}
  \psarc[linecolor=red,linewidth=2pt]{->}(3.75,-8){0.75}{0}{360}
  \rput[b](3.75,-7){$I_{\circlearrowleft}$}
\end{pspicture}
  \caption{Definition of the Stokes parameters from the measurement point of view according to the
\cite{Iau73} convention. The notation  $\Ia{\alpha}$ represents the intensity measured through
perfect linear polarizer with a  polarizing  direction making an angle $\alpha$ with respect to a
reference direction, taken to be   the  North--South, so that +Q
corresponds to a linear polarisation aligned with the North--South axis. $I_{\circlearrowright}$
and $I_{\circlearrowleft}$ represent  the intensity measured through perfect circular polarizer
selecting respectively right-handed and left-handed circular polarisation.}
  \label{fig:stokes-param}
  \end{center}
\end{figure}

\section*{Acknowledgements}
JM acknowledges the support of the Alexander von Humboldt foundation for his
research work in Göttingen. It is a pleasure to thank Thomas Gastine for his insightfull comments
about dynamo theory.

\end{document}